\documentclass{article}
\usepackage[utf8]{inputenc}

\pdfoutput=1
\usepackage[pdftex]{color}
\usepackage{amssymb}
\usepackage{amsthm}
\usepackage{amsmath}
\usepackage{latexsym}
\usepackage{amscd}
\usepackage{graphicx}
\usepackage[pdftex, colorlinks=true, citecolor=green]{hyperref}
\usepackage{lscape}
\usepackage{multirow}

\newcommand{\ds}{\displaystyle}

\newcommand{\ben}{\begin{equation}}     
\newcommand{\eeqn}{\end{equation}}
\newcommand{\bey}{\begin{eqnarray}}
\newcommand{\eey}{\end{eqnarray}}

\newtheorem{thm}{Theorem}[section]

\newtheorem{defn}[thm]{Definition}

\begin{document}

\vspace{5mm}
\noindent {\Large
\textbf{Synchronization and clustering in complex quadratic networks}
}
\\\\
\indent Anca R\v{a}dulescu$^{*,}\footnote{Associate Professor, Department of Mathematics, State University of New York at New Paltz; New York, USA; Phone: (845) 257-3532; Email: radulesa@newpaltz.edu}$, Danae Evans$^{2}$, Amani-Daisa Augustin$^3$, Anthony Cooper$^1$,\\
\indent Johan Nakuci$^4$, Sarah Muldoon$^5$
\\
\indent $^1$ Department of Mathematics, SUNY New Paltz
\\
\indent $^2$ Department of Physics,  SUNY New Paltz
\\
\indent $^3$ Department of Biology, SUNY New Paltz
\\
\indent $^4$ School of Psychology, Georgia Insitute of Technology
\\
\indent $^5$ Department of Mathematics, University at Buffalo

\begin{abstract}
\noindent In continuation of prior work, we investigate ties between a network's  connectivity and ensemble dynamics. This relationship is notoriously difficult to approach mathematically in natural, complex networks. In our work, we aim to understand it in a canonical framework, using complex quadratic node dynamics, coupled in networks which we call complex quadratic networks (CQNs). 

After previously defining extensions of the Mandelbrot and Julia sets for networks, we currently focus on the behavior of the node-wise projections of these sets, and on defining and analyzing the phenomena of node clustering and synchronization. We investigate the mechanisms that lead to nodes exhibiting identical or different Mandelbrot set. We propose that clustering is strongly determined by the network connectivity patterns, with the geometry of these clusters further controlled by the connection weights. We then illustrate the concept of synchronization in an existing set of whole brain, tractography-based networks obtained from 197 human subjects using diffusion tensor imaging.

Synchronization and clustering are well-studied in the context of networks of oscillators, such as neural networks. Understanding the similarities to how these concepts apply to CQNs contributes to our understanding of universal principles in dynamic networks, and may help extend theoretical results to natural, complex systems.
\end{abstract}

\section{Introduction}

\subsection{Background on dynamic networks}

Understanding dynamics in coupled networks has become recently an important research direction in a variety of fields. Mathematically, it is a rich topic, that generates interesting questions at the intersection of graph theory and dynamical systems, the complexity of which depends equally on the type of node-wise dynamics, and on the network size and architecture. Mathematical inquiries and results on dynamic networks are relevant to many questions in the life sciences, since many natural systems are organized as complex, adaptable networks. Understanding how the hard-wiring of these systems affects their function is a crucial, structural question, whether the system in question is the brain, a social or an epidemic network. 

An important direction of study in dynamic networks relates to their long-term ensemble behavior, and the dependence of this behavior on the system's parameters. These parameters may characterize the individual dynamics of the network's nodes or the connectivity between them, including both connection distribution and connection strengths. For example, if the system describes firing rates in the neural populations that form a brain circuit, one aspect crucial to understand is whether the system converges in the long term to sustained firing (stable equilibrium), or to firing rate oscillations (stable limit cycle). This is important from an application standpoint, since these two different asymptotic patterns reflect into different functional and cognitive outcomes for the circuit in question. Another aspect which was proved to be crucial to a circuit's functional dynamics is that of synchronization in the nodes' activity. Neural populations have been observed to act completely synchronously or asynchronously, or to separate into clusters with different activity patters. On one hand, synchronization and clustering are highly dependent on the network interconnectivity and architecture; on the other hand, at a functional level, they have been shown to reflect onto the type and efficiency of network function to be performed.

These different possible outcomes in dynamic activity of a network depend on intrinsic regulatory parameters that govern the behavior of each unit, in absence of the coupling (node-wise dynamics); but they also depend on the coupling itself: which nodes are connected (e.g. via synaptic pathways in a neural network), and how strong these pathways are in time. Studying this relationship between the network structure and the emergent coupled dynamics is central to the study of dynamic networks. For example, in neuroscience, some psychiatric disorders have been correlated both with atypical brain connectivity patters between specific brain areas, as well as with abnormal neural rhythms in certain these areas. It is important to understand how the former contribute to the latter, and how this combination further leads to a certain ensemble behavior in the brain network versus another.

This tie between a network's hard-wiring and function has been studied intensely for the past decade (e.g., in neuroscience, through the Human Connectome Project). It has become increasingly clear that establishing a precise correspondence between connectivity and emergent dynamic patterns is a very difficult problem both analytically and computationally, especially when considering the sheer size and complexity of natural networks, with the brain in particular. Newer research has instead focused on understanding the principles of this correspondence in smaller, simpler networks (in terms of network size and architecture, or in terms of node-wise dynamics, or both).

\subsection{Complex quadratic networks, as an analytic framework}

In our previous work, we adopted the simplified framework of networked logistic maps as a starting point for approaching basic dynamic questions in the context of networks~\cite{radulescu2017real,radulescu2019asymptotic}. That is, we considered nodes with complex quadratic dynamics in the family $f_c \colon \mathbb{C} \to \mathbb{C}$, $f_c(z) = z^2+c$. More specifically, in this framework, each network node receives weighted inputs from the adjacent nodes, and integrates these inputs in discrete time as a complex quadratic map. Then the system takes the form of an iteration in $\mathbb{C}^n$:

\begin{eqnarray}
z_j(t) \longrightarrow z_j (t+1) &=& f_j\left(\sum_{k=1}^{n}{g_{jk} A_{jk} z_k} \right) \nonumber 
\label{mothermap}
\end{eqnarray}

\noindent where $n$ is the size of the network, $\ds A=(A_{jk})_{j,k =1}^n$ is the binary adjacency matrix of the oriented underlying graph, that is $A_{jk}=1$ if there is an edge from the node $k$ to the node $j$, and $A_{jk}=0$ otherwise. The coefficients $g_{jk}$ are the signed weights along the adjacency edges (in particular, $g_{jk}=0$ if there is no edge connecting $k$ to $j$, that is if $A_{jk}=0$). In isolation, each node $z_j(t) \to z_j(t+1)$, $1 \leq j \leq n$, iterates as a quadratic function $f_j(z) = z^2 + c_j$. When coupled as a network with adjacency $A$, each node will act as a quadratic modulation on the sum of the inputs received along the incoming edges (as specified by the values of $A_{jk}$, for $1 \leq k \leq n$). 

This is a simplified framework, which turned out to preserve, often in a weaker form, some of the properties and results determined in the traditional case of a single iterated quadratic map (e.g., the existence of an escape radius is guaranteed in some types of networks, but may fail in others). We used this to our advantage, and proceeded to study theoretically the effects of network architecture on its long-term dynamics. We have been generally investigating whether one can use properties of multi-dimensional orbits in $\mathbb{C}^n$, in particular their asymptotic behavior (via the topological and fractal structure of Julia and Mandelbrot multi-sets), to classify dynamic behavior for different network architectures. To this aim, we defined network asymptotic sets as follows:


\begin{defn}
We define the \textbf{multi-Mandelbrot set (or the multi-M set)} of the network the parameter locus of $(c_1,...,c_n) \in \mathbb{C}^n$ for which the multi-orbit of the critical point $(0,...,0)$ is bounded in $\mathbb{C}^n$. We call the \textbf{equi-Mandelbrot set (or the equi-M set)} of the network, the locus of $c \in \mathbb{C}$ for which the critical multi-orbit is bounded for \textbf{equi-parameter} $(c_1,c_2,...c_n)=(c,c,...c) \in \mathbb{C}^n$.  
\end{defn}

In our previous work, we focused on describing topological properties of these asymptotic, global network sets. We aimed to understand which geometric aspects survive when generalizing these objects from the context of a single iterated map to the richer dynamics of ensemble iterations of a whole network.The difficulty comes from the fact that, even in the situation where the nodes are identical, the connectome favors nodes differently -- hence the strong dependence of the topology of the asymptotic sets on the architecture of the network. We have been therefore trying to track down how the connectivity profile of the network affects the shapes, properties and relationships between asymptotic sets.

For example, note that the Mandelbrot set of an $n$-dimensional network was defined as the node parameter range for which the critical point (i.e., all nodes equal zero) is bounded in $\mathbb{C}^n$ under ensemble iterations of the network. In the traditional case of single iterated quadratic maps, this is equivalent to Julia set connectedness locus. One of our central questions in previous work has been whether this result remains true for networks. Numerical experiments strongly suggest that this equivalence no longer holds in the case of iterating networks (for either Julia sets, or uni-Julia sets). In our previous work, we conjectured a weaker form of this result, for the case of identical nodes. We suspect (but have not yet proven analytically) that connected uni-J sets only occur for $c$ in the network equi-M set, and that totally disconnected uni-J imply a $c$ outside of the equi-M set. These seems to be a transitional region $c \in \mathbb{C}$ for which the uni-J set is disconnected, but not totally disconncted; this region contains the border of the network's equi-M set. 

This is not surprising, since in order for a network's critical orbit to be bounded, all of its node-wise components need to be simultaneously bounded, a complex condition that depends crucially on the shape of the network. A natural question that emerges is that of how different can node-wise behaviors be, between different nodes of the same network. We have shown before that even in networks with identical node dynamics (i.e., all nodes have the same parameter $c$) the nodes may exhibit different asymptotic behaviors (in the sense of their own, node-wise, Mandelbrot and Julia sets). Node coupling determines the communication between two or more nodes, and different types of networking motifs seem to lead to network Mandelbrot and Julia sets of different shapes and sizes. In the current paper, our interest is focused on understanding node ``synchronization.'' We aim to analyze and interpret the distinct effects of varying connection patterns on synchronization. We are also interested to understand the relationship between node de-synchronization, and failure of the Fatou-Julia theorem.

\section{Node-wise synchronization and clusters in CQNs}
\subsection{Definitions}
\label{definitions}

In all the following sections we will be working with $n$-dimensional networks with complex nodes $z_1 \cdots z_n$ with identical node-wise dynamics (i.e., all nodes are governed by the same complex parameter $c$). 

\begin{defn}
For each node $z_k$, $1 \leq k \leq n$, we define the \textbf{node-wise equi-M set} ${\cal M}_k$ as the parameter locus $c \in \mathbb{C}$ for which the component corresponding to the node $z_k$ of the critical multi-orbit is bounded. 
\end{defn}

\noindent Note that, with this definition, the network equi-M set is the intersection of all node-wise equi-M sets. Part of our mission in this paper is to understand what conditions may determine whether the equi-M sets for two nodes are identical or different. To help with this aim, we introduce some additional terminology:

\begin{defn}
We say that two network nodes $z_i$ and $z_j$ are \textbf{M-synchronized}, if their equi-M sets ${\cal M}_i$ and ${\cal M}_j$ are identical. We also say that the nodes belong to the same \textbf{M-synchronization cluster} (or simply \textbf{M-cluster}) of the network.
\end{defn}

\begin{figure}[h!]
\begin{center}
\includegraphics[width=0.8\textwidth]{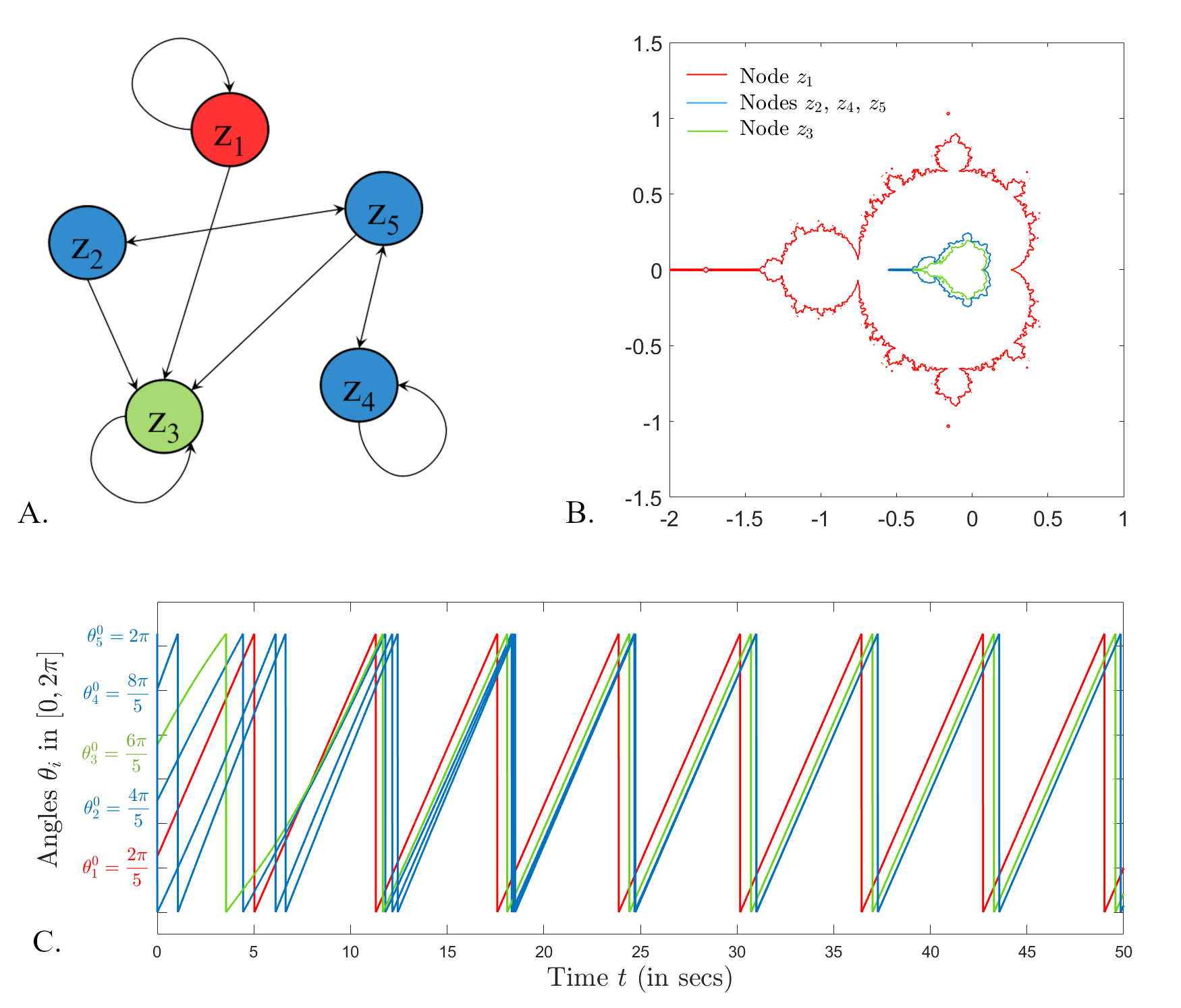}
\end{center}
\caption{\emph{\small {\bf Example of M-clusters in a 5-dimensional network.} {\bf A.} Graph representation of the connectivity matrix of our example network. {\bf B.} Illustration of the three different M-clusters formed by the CQN: red cluster (node $z_1$); blue cluster (nodes $z_2$, $z_4$ and $z_5$) and green cluster (node $z_3$). {\bf C.} Illustration of the same three different synchronization clusters formed by the Kuramoto model with parameters $w=1$ and $K=1$, and balanced initial conditions around the circle.}}
\label{example_5dim_M_clusters}
\end{figure}

\noindent Figure~\ref{example_5dim_M_clusters} illustrates the concept of M-synchronization for an example network with $n=5$ nodes, with connectivity scheme illustrated in the first panel of the figure, corresponding to the adjancency matrix:
\begin{equation}
A = \left( \begin{array}{ccccc} 
1 & 0 & 0 & 0 & 0\\
0 & 0 & 0 & 0 & 1\\
1 & 1 & 1 & 0 & 1\\
0 & 0 & 0 & 1 & 1\\ 
0 & 1 & 0 & 1 & 0
\end{array} \right)
\label{5Dmatrix}
\end{equation}

\noindent and all unit weights $g_{ij}$ on the nonzero edges (i.e., the adjacency matrix is also the connectivity matrix). In other words, the corresponding 5-dimensional discrete iteration is given by following scheme (where all nodes update simultaneously):
\begin{eqnarray*}
z_1 &\to& z_1^2+c\\
z_2 &\to& z_5^2+c\\
z_3 &\to& (z_1+z_2+z_3+z_5)^2+c\\
z_4 &\to& (z_4+z_5)^2+c\\
z_5 &\to& (z_2+z_4)^2+c
\end{eqnarray*}

The five nodes separate into three different M-clusters, the M sets of which are shown as contours in Figure~\ref{example_5dim_M_clusters}b: $z_1$, which is only driven by self-input, generates a node-wise M set identical with the traditional Mandelbrot set (the boundary of which is shown in red). Nodes $z_2$, $z_4$ and $z_5$ are M-synchronized into one M-cluster; their identical M sets are sketched with a blue contour. Node $z_3$ forms the third M-cluster, the set of which is shown as a green contour. 

The concept of synchronization has been widely studied in networks of continuous-time dynamic oscillators, with brain networks in particular. In this more traditional form, synchronized activity of two nodes is understood as convergence to an identical temporal oscillatory pattern. To better understand how our concept of M-synchronization relates with the standard definition, we illustrate the latter in the context of the Kuramoto model  (as an example of a simple, generic model of network dynamics). Synchronization patterns in Kuramoto coupled dynamics, and how they are actualized in the brain connectome have been studied over the past decade~\cite{odor2019critical}). For similarity with CQNs with identical nodes, we consider the Kuramoto model with $N$ identical coupled oscillators

\begin{equation}
\frac{d\theta_i}{dt} = w + \frac{K}{N} \sum_{j=1}^{N} A(i,j) \sin(\theta_j-\theta_i)
\label{Kuramoto_net}
\end{equation}

\noindent where the variables $\theta_i$ can be viewed as angles modulo $2\pi$. Here, $w$ is the common internal frequency, $K$ is the control parameter, and $A$ represents (similarly to our model) the connectivity weight matrix. Depending on the context of $w$, $K$ and $A$, subsets of nodes in the network may converge in time to identical or out of phase oscillations in $[0,2\pi]$, creating complex synchronization patterns. The $N$-dimensional network can be then subdivided into subsets (or clusters) of temporally synchronized nodes, ranging from one cluster (in the case where all nodes are eventually synchronized) to $N$ distinct clusters (when all nodes perform, in the long term, distinct/out of phase oscillations). As an initial comparison, we show in Figure~\ref{example_5dim_M_clusters}c an example of clustering in the Kuramoto network with the same connectivity matrix as the one used for the M-clusters. In this particular example, the activity in the Kuramoto network readily synchronizes into the same three clusters as the CQN, as shown. However, this cannot be expected to happen in general; in fact for smaller values of the coupling $K$, the system may undergo other transients along the way to this asymptotic pattern, making it hard to even distinguish when the oscillations have settled around the stable periodic attractor. We will return to this phenomenon and its significance later in the paper.

\subsection{Effects of networking on M-clusters}
\label{effects}

We have previously shown that topological properties of the network equi-M set in CQNs depend strongly on the network connectivity patterns. For example, we have illustrated that, even in low-dimensional networks, small perturbations in connectivity (understood as small local changes in either position of the connecting edges, or in the connectivity weights) may lead to large differences in the size, shape and connectedness of the equi-M set. Here, we wish to establish that these differences stem in fact from the clustering behavior.

\begin{figure}[h!]
\begin{center}
 \includegraphics[width=0.8\textwidth]{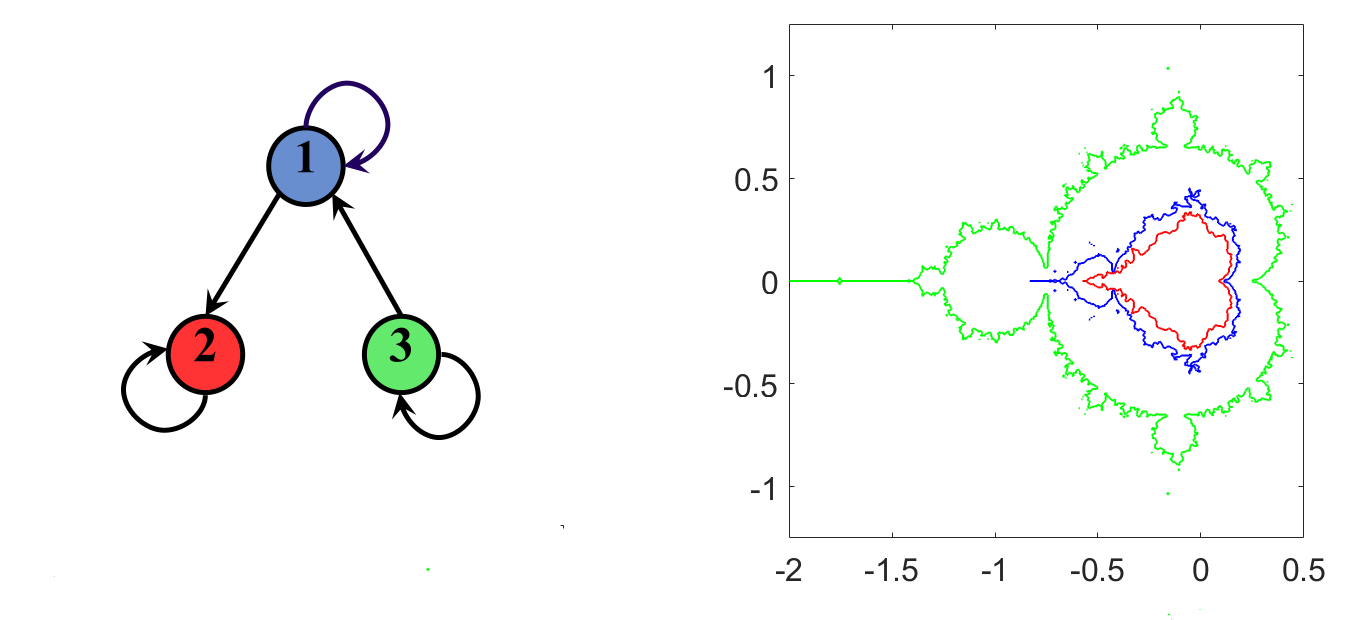}
\end{center}
\caption{\emph{\small {\bf Example of networks with nested equi-M sets.} {\bf Left.} Three-dimesional network defined by $z_1 (z_1+z_2)^2+c$, $z_2 \to (z_2+z_3)^2+c$, $z_3 \to z_3^2+c$. {\bf Right.} Corresponding contours of the node-wise equi-M sets (bottom). The colors of the contours are matched with those of the corresponding network nodes.}}
\label{examples}
\end{figure}

To begin with, it is easy to see from our definitions that the equi-M set of a network, is the (finite) intersection of all its node-wise Mandelbrot sets. This relationship may be even more meaningful than seems at first. For example, it appears from our simulations that, for many networks, this is a nested intersection. Below, we describe a 3-dimensional example, which we will also revisit later in the paper.\\

\noindent {\bf Example 1.} \emph{In the network defined in Figure~\ref{examples}, the nodes define three distinct, nested equi-M sets.}

\proof{We show first that, if the critical orbit is bounded in the $z_1$ component, it is also bounded in the $z_3$ component. Indeed, for any iteration $n\in \mathbb{N}$, we have the following:
\begin{eqnarray*}
\lvert z_1(n+1) \rvert = \lvert (z_1(n)+z_3(n))^2+c \rvert \geq \lvert z_1(n)+z_3(n)\rvert^2 - \lvert c \rvert
\end{eqnarray*}
Hence
\begin{eqnarray*}
\sqrt{\lvert z_1(n+1) \rvert + \lvert c \rvert } \geq \lvert z_1(n) + z_3(n) \rvert \geq \lvert z_3(n) \rvert - \lvert z_1(n) \rvert
\end{eqnarray*}
and thus
\begin{eqnarray*}
\lvert z_3(n) \rvert \leq \sqrt{\lvert z_1(n+1) \rvert + \lvert c \rvert } + \lvert z_1(n) \rvert
\end{eqnarray*}

\noindent Since $z_1$ is bounded, it follows that $z_3$ is also bounded. Consider now the critical orbit of the network under the parameter $c=-2$. It is easy to show that this critical orbit is eventually fixed in its $z_3$ component ($z_3(n) = 2$, for all $n \geq 2$), hence $-2 \in {\cal M}(z_3)$. We will show that the $z_1$ component of the critical orbit escapes. Notice first that the sequence $z_1(n)$ has all positive real values, and that, for $n \geq 2$
\begin{eqnarray*}
z_1(n+1) \leq (z_1(n) + 2)^2 -2 = z_1^2(n)+4z_1(n)+2 > 2z_1(n)
\end{eqnarray*}

\noindent Since $\ds \frac{z_1(n+1)}{z_1(n)} >2$ for $n \geq 2$, it follows that $z_1(n) \to \infty$. This proves that ${\cal M}(z_1) \subsetneq {\cal M}(z_3)$. For the other connected node pair, one can show that ${\cal M}(z_2) \subsetneq {\cal M}(z_1)$ very similarly to the inclusion above. In addition, $c=-3/4 \in {\cal M}(z_2)$, but $3/4 \not\in {\cal M}(z_1)$. To show this, notice that the $z_3$ component of the critical orbit is eventually fixed to $-3/4$, and the $z_2$ component of the critical orbit eventually oscillates between $-3/4$ and $3/2$. Hence the $z_1$ component is a sequence of positive real numbers, which builds up recursively, for every even $n \in \mathbb{N}$, as:
\begin{eqnarray*}
z_1(n+2) &=& \left([(z_1(n)+z_2(n))^2+c]+z_2(n+1) \right)^2+c = (z_1(n)^2+3z_1(n)+3/4)^2-3/4
\end{eqnarray*}
\noindent which clearly diverges. \qed}\\



The nesting property does not hold in general. One can simply think of the trivial example in which the nodes are independent, with different levels of self-input: $z_1 \to (a_1z_1)^2+c$, $z_2 \to (a_2z_2)^2+c$, $z_3 \to (a_3z_3)^2+c$, $a_1<a_2<a_3$. Notice that the components of the critical orbit scale linearly with the value of the parameter; hence the node-wise M sets also scale: if $c \in {\cal M}(z_1)$, then $\frac{a_1}{a_2}c \in {\cal M}(z_2)$. Then, for values of $a_1$ and $a_2$ close enough together, the sets intersect nontrivially.

There are, however, more complex examples of networks for which there exist two nodes with M sets intersecting non-trivially, as shown below.
Figure~\ref{counterexample} illustrates the geometry of the M set clusters for the $N=20$ dimensional network with adjacency matrix A1 (shown in Appendix B), and unit edge weights. The network exhibits eight distinct M clusters (shown on the left), at least two of which (illustrated separately on the right, for clarity) intersect non-trivially. In future work, we are interested to further investigate necessary and sufficient network conditions for nesting of node-wise uni-J and M sets.

\begin{figure}[h!]
\begin{center}
\includegraphics[width=0.8\textwidth]{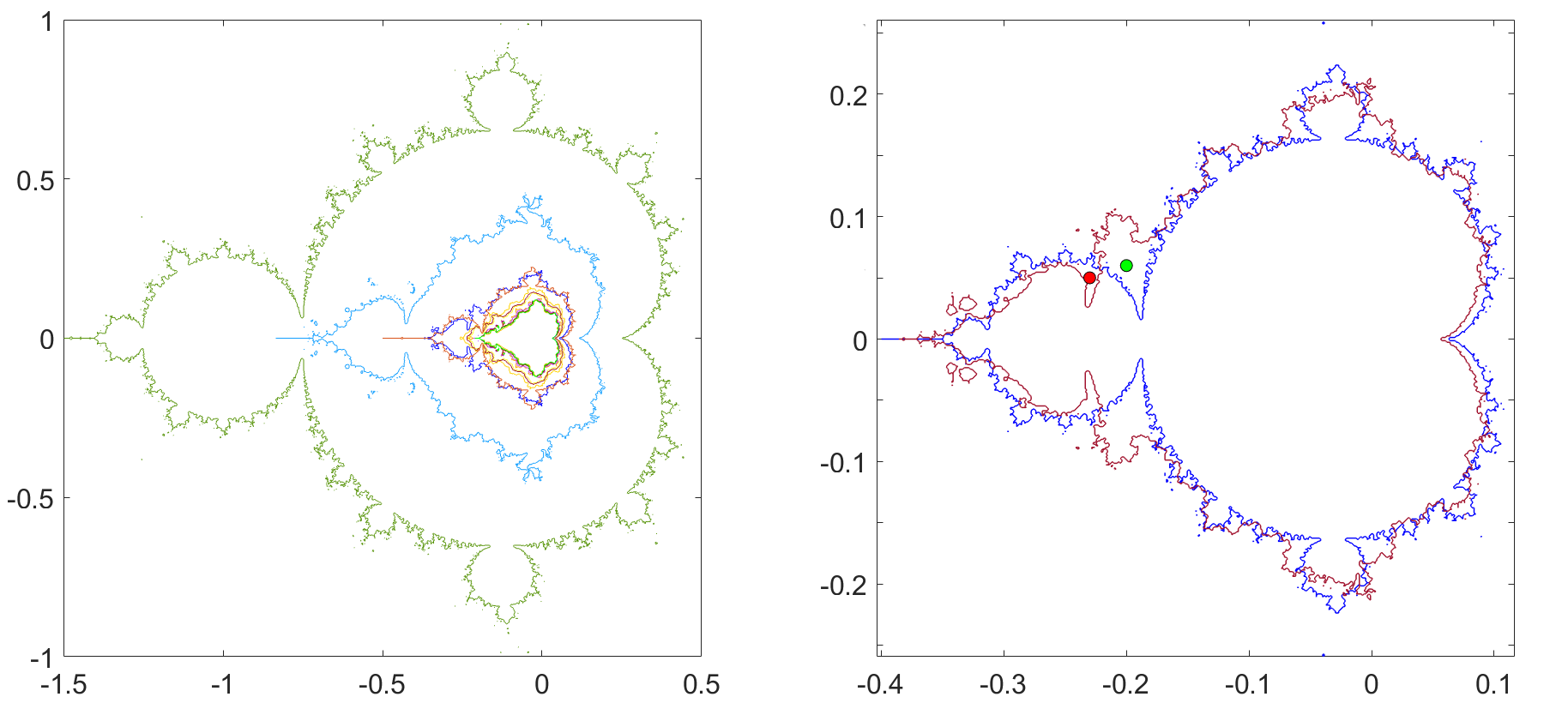}
\end{center}
\caption{\emph{\small {\bf Example of network with non-trivial M set intersections.} {\bf Left.} In the 20-node network with adjacency matrix $A_1$ (shown explicitly in Appendix B) and unit edge weights, the nodes synchronize into eight distinct M clusters (containing from one to four nodes each). Their contours are illustrated in different colors. The contours are not all nested, i.e. some of them intersect non-trivially. {\bf Right.} The panel shows in more detail two of the M cluster contours on the left which have a non-trivial intersection; we specifically marked one point ($c$=$-0.23+0.05i$, in red) which is in the blue cluster, but not in the maroon cluster, and one point ($c$=$-0.2+0.06i$, in green) which is in the maroon cluster, but not in the blue one. The computations supporting this panel can be found in Appendix A.}}
\label{counterexample}
\end{figure}

Additional robust properties of M sets have been previously pointed out, such as the persistence of the cusp, despite the fact that network M sets don't otherwise preserve the hyperbolic bulb structure of the traditional Mandelbrot set. While the shape and structure of the tail varies wildly with that of the network (including the potential to break off into separate connected components), the cusp is a common geometric feature in every node-wise M set. 

In this context, it becomes clear that, to understand how the geometry of the network equi-M set depends on connectivity patterns, it is helpful to first understand how the number and shape of the node-wise clusters depend on these patterns. In particular, we are interested to study the differences between the effects on clustering of perturbing global connectivity (number of edges), versus local connectivity patterns (removing, adding or replacing one edge), and versus altering the connection strengths (increasing and decreasing one or multiple edge weights).\\

\noindent \emph{\textbf{Local changes in edge placement and weights.}} Our simulations suggest that simply strengthening or weakening one edge can alter the shape of the node-wise M-sets in one or more clusters, while leaving the cluster separation unchanged. In contrast, adding or removing one edge completely can promptly change the cluster structure, in addition to potential changes in the shape of the sets corresponding to each cluster. 

\begin{figure}[h!]
\begin{center}
\includegraphics[width=0.9\textwidth]{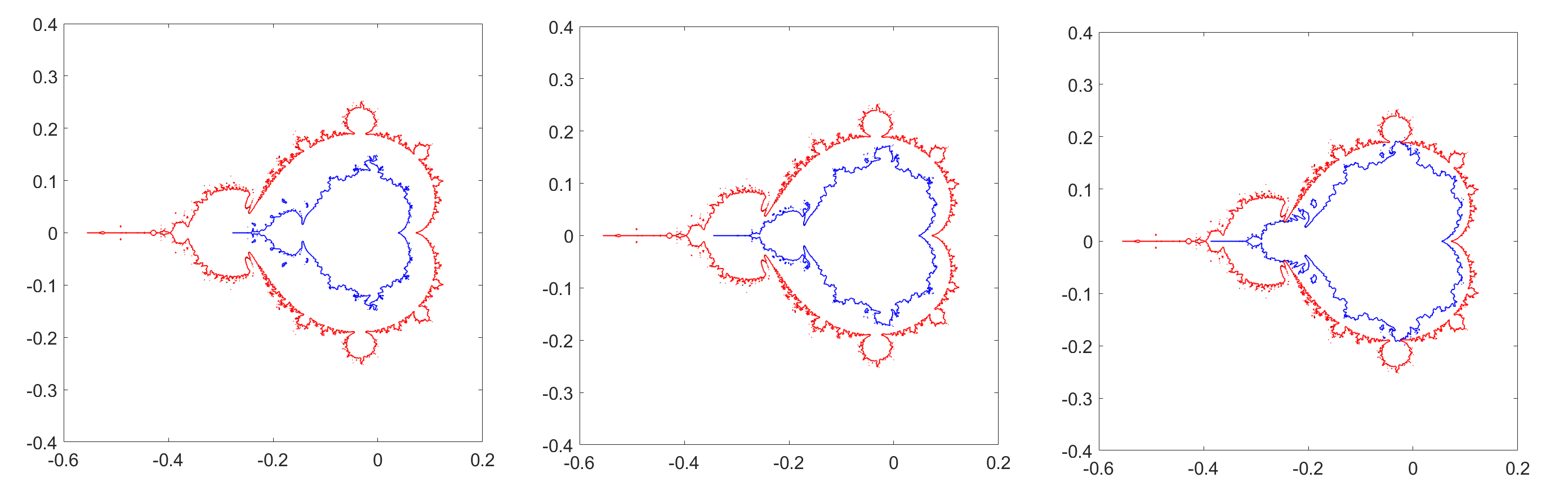}
\end{center}
\caption{\emph{\small {\bf Effect of edge weight on equi-M set clustering and shape.} Three different levels of cross-talk in the network lead to the same dynamic clustering: cluster 1, in blue (node 1 and 3); cluster 2, in red (nodes 2, 4, 5); however, the shape of cluster 2 depends non-trivially on the cross-talk level $A_{12}$. {\bf A.} $A_{12}=1$; {\bf B.} $A_{12}=0.5$; {\bf C.} $A_{12}=0.01$.}}
\label{5dim_edge_weight}
\end{figure}

To illustrate, consider the 5-dimensional matrix obtained from the connectivity matrix $A$ described in example~\eqref{5Dmatrix}, to which we add cross-talk from the node $z_3$ to the node $z_1$. The panels of Figure~\ref{5dim_edge_weight} show how the shape of the clusters changes when the weight of this new edge decreases from $A_{13}=1$ to $A_{13} \to 0$. Notice that, throughout this process, the two-cluster structure remains unaltered, although the shape of one of the clusters is affected by the perturbations in the local weight. As $A_{13}$ becomes zero, a bifurcation occurs in the clustering structure, to recover the original 3-cluster formula in Figure~\ref{example_5dim_J_clusters} (due to the fact that node $z_1$ has now been decoupled from any input coming from other network nodes and is only being driven by the self-loop, hence the associated M set is the traditional Mandelbrot set).

\begin{figure}[h!]
\begin{center}
\includegraphics[width=0.9\textwidth]{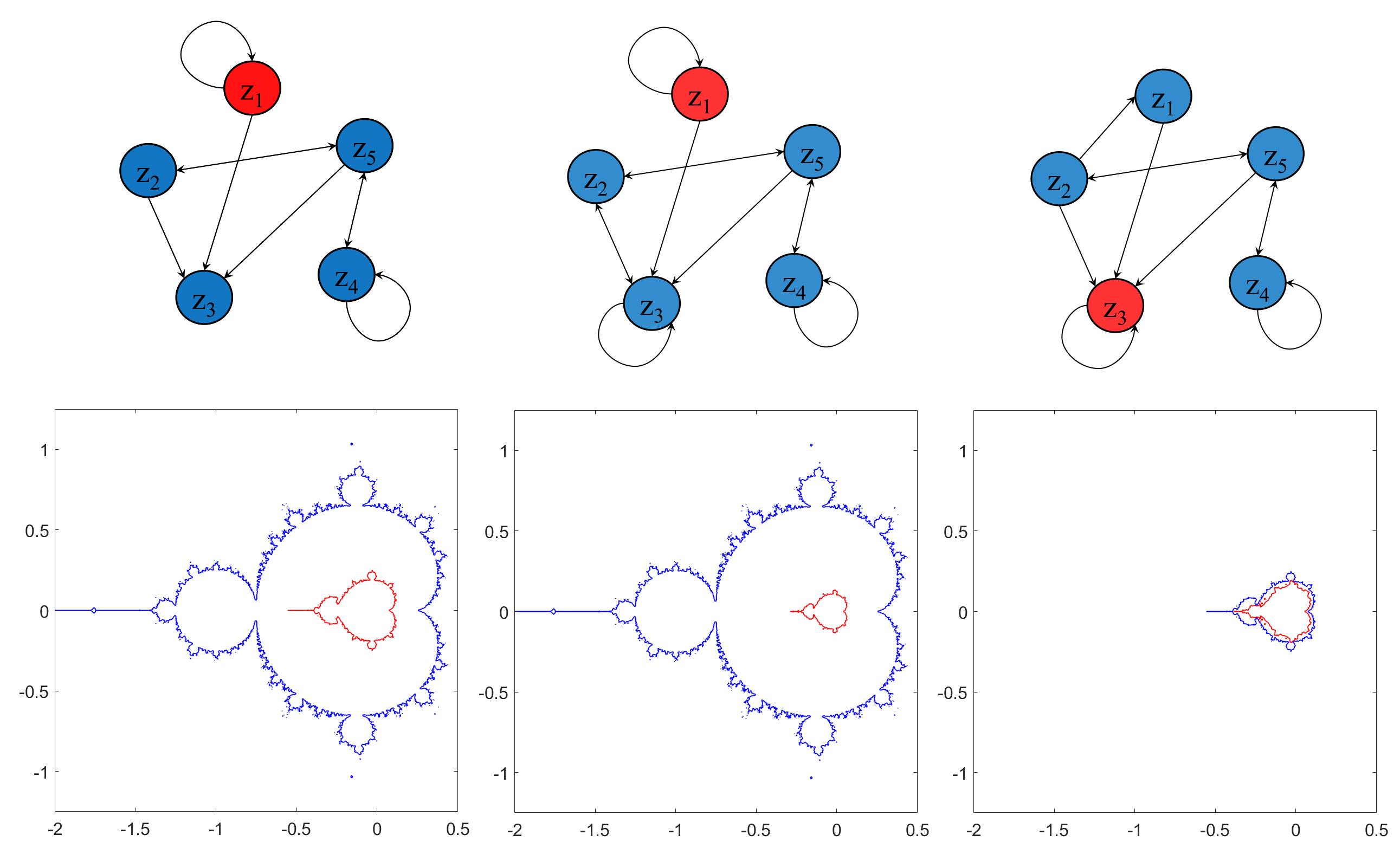}
\end{center}
\caption{\emph{\small {\bf Effect of edge position on equi-M set clustering and shape.} {\bf A.} Deleting the input $A_{32}=1$ leads to two clusters: cluster 1 (node 1, in blue) and cluster 2 (all other nodes, in red). {\bf B.} Adding $A_{33}=1$ leads to the same two clusters. {\bf C.} Deleting the input $A_{11}=1$ and adding the input $A_{12}=1$ leads to two different clusters: cluster 1 (nodes 1,2,4,5, shown in blue) and cluster 2 (node 3, shown in red). }}
\label{5dim_M_clusters}
\end{figure}

In Figure~\ref{5dim_M_clusters}, we continue by illustrating how local connectivity changes (consisting of removing or adding one edge), applied to the matrix $A$, affect clustering and the shape of the M set in each node. For example, removing the self-loop on node $z_3$ (i.e., setting $A_{33}=0$) allows the node to synchronize with the others in the second cluster for the original matrix $A$, and the resulting network will now only have two clusters (panel A). The shape of the M set for this enlarged cluster is different, however, than the shapes of either the two merging clusters for the original matrix $A$. Adding an equal weight cross-talk from node $z_3$ to $z_2$ (i.e. setting $A_{21}=1$) has the same effect of synchronizing nodes $z_2-z_5$, into a cluster with a new shape for the M set. Finally, replacing the self-input on node $z_1$ by a same weight input from node $z_2$ synchronizes $z_1$ with the rest of the nodes in cluster 2 of $A$, without changing the shape of the remaining clusters.\\

\begin{landscape}
\begin{figure}[h!]
\begin{center}
\includegraphics[width=\linewidth]{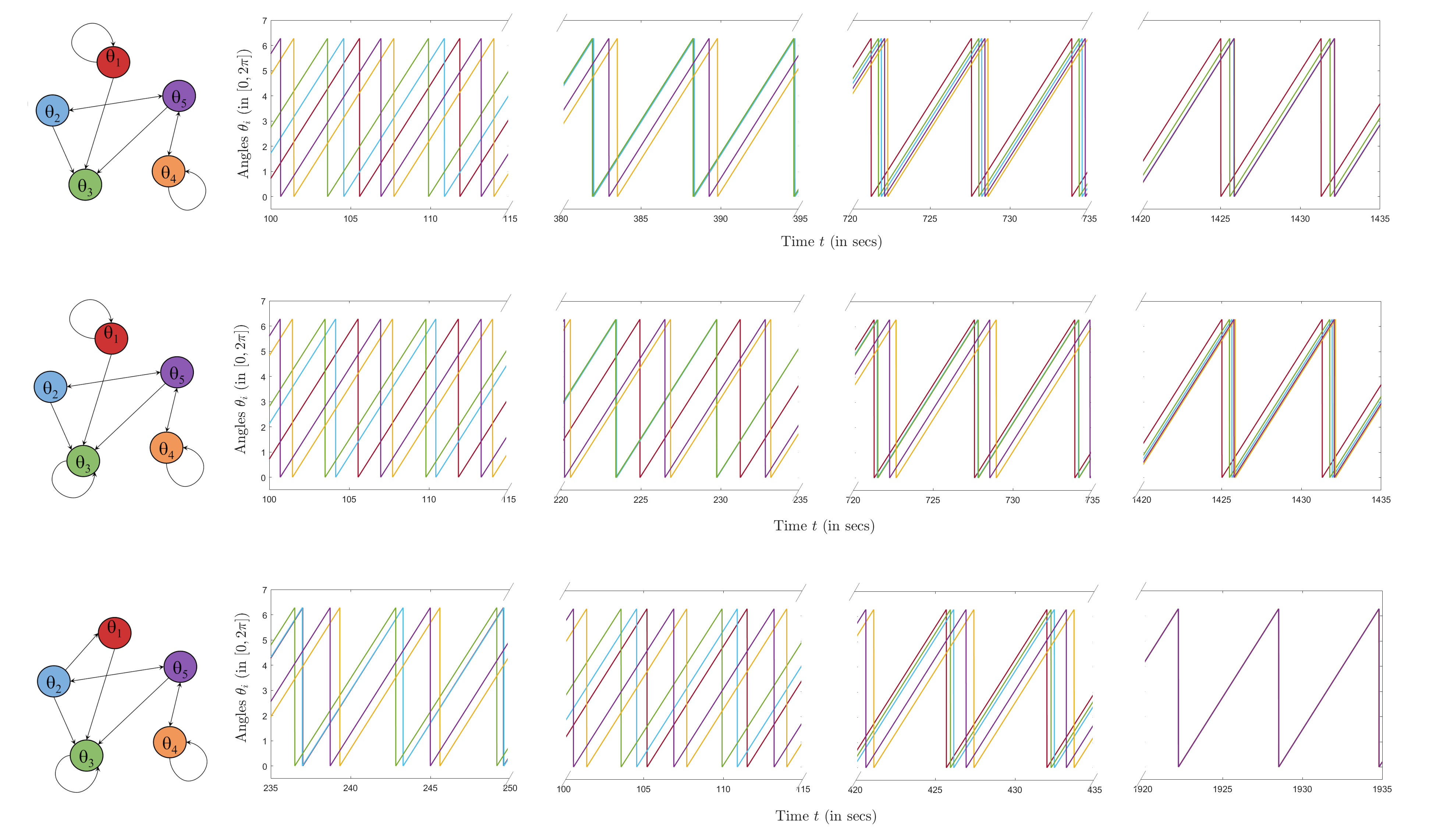}
\end{center}
\caption{\emph{\small {\bf Effect of edge position on synchronization and clustering in the Kuramoto model.} The three panels show the evolutions of the solutions to the Kuramoto model with $N=5$ coupled nodes, $w=1$, $K=0.02$ and balanced initial conditions $\ds \theta_i^0 = \frac{2i\pi}{5}$, for $1 \leq i \leq 5$, as the connectivity matrix undergoes the same changes as those described in Figure~\ref{5dim_M_clusters}. In order to better emphasize how synchronization patterns change over time, we selected to illustrate specific time windows, separated by grey shaded blocks.}}
\label{5dim_Kuramoto_clusters}
\end{figure}
\end{landscape}

Continuing the comparison in Section~\ref{definitions}, Figure~\ref{5dim_Kuramoto_clusters} shows the synchronization clusters formed in the Kuramoto model with five identical nodes and the same connectivity matrices as in Figure~\ref{all3D}. The clustering patterns in the Kuramoto model depend on the initial conditions and on how these fall into the attraction basins of various stable oscillations. In particular, identical initial conditions produce trivial clustering behavior (nodes immediately synchronize to one cluster, independently on the network). By analogy with the criticality condition in equi-M sets, we considered in for our simulation in Figure~\ref{5dim_Kuramoto_clusters} initial conditions with balanced phases around the circle, for which the order parameter is zero. Even with fixed initial states, the temporal evolution of the Kuramoto system is very complex, and quite difficult to interpret. Even for the simple networks at hand, the system exhibits chimeras, with synchronization patterns evolving in time, making it difficult to pin down asymptotic states, and even more difficult to compare the system's long-term behavior between different connectivity schemes. Figure~\ref{5dim_Kuramoto_clusters} illustrates this comparison for the same three matrices used in Figure~\ref{all3D} for CQNs. Notice that, since they start with the same initial conditions, the oscillations are rather similar between the three examples for the first transient segment (panels 1). However, these oscillations weave into different evolving synchronization patterns, according to the underlying matrix. For the first connectivity scheme $A$, nodes 3,4 and 5 initially synchronize (panel 2), then de-synchronize (panel 2), then nodes 1,2 and 4 synchronize, showing three clusters at the end of our simulation. Even though the final result is similar for the second connectivity scheme (three clusters and the same grouping as above), this system crosses a 4-cluster transient. Comparing the two systems prematurely in time may suggest different behaviors, when at the end of the simulation they are very similar. Conversely, our third example shows four transient clusters around t=250, same as example 2, but its asymptotic behavior is completely different (all nodes fully synchronize by the end of the simulation run $t=1450$. In conclusion, while it is clear that the connectivity matrix $A$ controls to a large extent the Kuramoto patterns, the assessment of how and to what extent this is the case relies on choosing the appropriate moment in time to make this assessment, which is an additional complication to an already complex problem. When working with CQN networks, a good approximation of asymptotic behavior is guaranteed in the case of equi-Mandelbrot sets by escape radius theorems, eliminating in most cases the question of whether the set would further change under prolonged iterations.

How the behavior of the balanced initial conditions is relevant to synchronization in the Kuramoto model for more general solutions is a well-studied area, beyond the scope of our paper. However, the analogous question in CQNs, asking to what extent the critical orbit truly encapsulates the behavior of all possible network orbits, is a new and important extension of traditional results in single quadratic map iterations, and is further discussed in Sections~\ref{connections}. A more careful comparison of parameter and network dependence between the two models (CQN and Kuramoto) is carried out in Section~\ref{Kuramoto}.\\

\begin{figure}[h!]
\begin{center}
\includegraphics[width=0.9\textwidth]{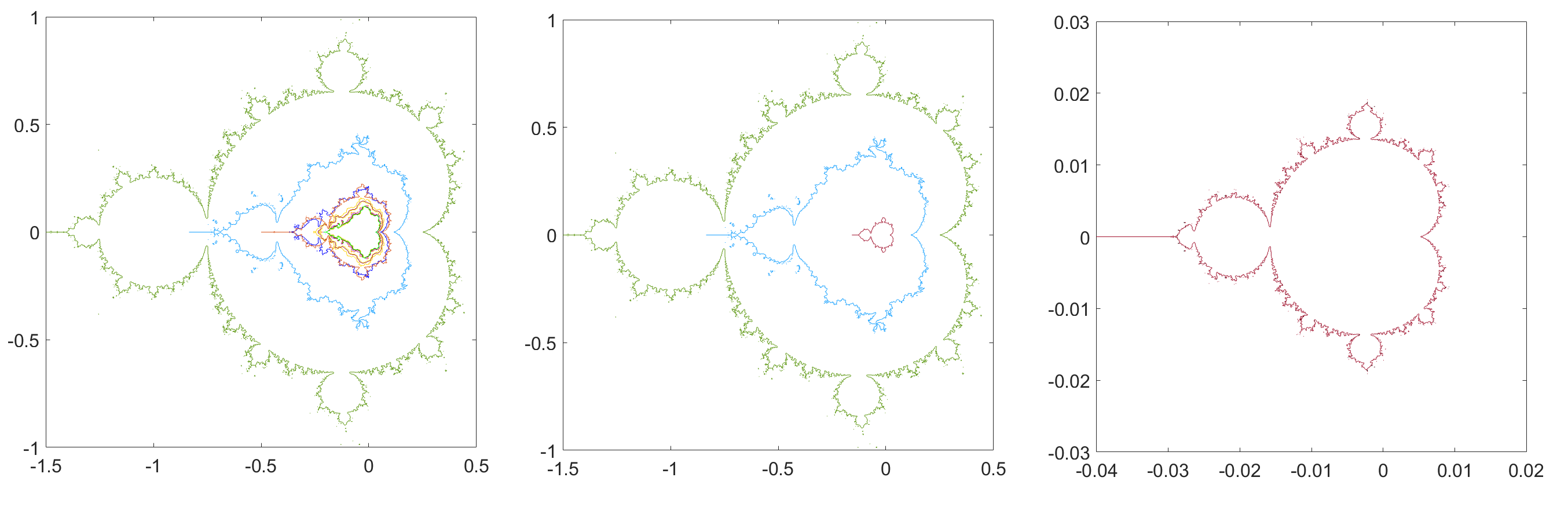}
\end{center}
\caption{\emph{\small {\bf Effect of edge density on equi-M set clustering and shape.} {\bf A.} M clusters for a networks with $N=20$ nodes and $\delta = 50$ edges (20 of which are self-loops). The contours corresponding to the eight different clusters are shown in different colors. Notice that at least two contours intersect (hence in this case the M sets are not nested). {\bf B.} M set clusters for a networks with $N=20$ nodes and $\delta = 118$ edges (20 of which are self-loops). The contours corresponding to the three different clusters are shown in different colors. In this case, the M sets are nested. {\bf B.} M set cluster for a networks with $N=20$ nodes and $\delta = 133$ edges (20 of which are self-loops). In this case, all M sets are synchronized. (Also notice the smaller size of the set in this case.) In all three cases, all edges have unit weights. The specific connectivity matrices that were used are shown in Appendix B.}}
\label{20dim_M_clusters}
\end{figure}

\noindent \emph{\textbf{Global changes to edge density.}} On a more global scale, our simulations suggest that the number of edges correlates with both the size of the M sets and their clustering, in the sense that a higher number of edges leads to smaller and more synchronized node-wise M sets. This is not surprising, and has been suggested in our prior work. We illustrate this relationship further in Figure~\ref{20dim_M_clusters}, in which we consider the behavior in three different networks with $N=20$ nodes, unit edge weights, and progressively higher edge densities (from panel a to panel c). The network in panel a shows eight different clusters, with at least two M sets intersecting non-trivially. The network in panel b exhibits three nested M sets, and panel c illustrates a network where all M sets are synchronized. This follows the intuition that, the denser the matrix -- the more synchronized the M sets. We will reiterate this idea when illustrating synchronization with an example from brain connectomics, in Section~\ref{connectomics}.

Of course, a more comprehensive analysis needs to be performed to establish whether these effects remain significant across all possible matrices with these properties. For the start, this is more easily done for lower dimensional network spaces, where there are less configurations to investigate. In the next section, we consider all 3-dimensional oriented networks with self loops, and fixed edge weights (=1), and we study more carefully the effect that the re-positioning of any subset of edges produces on the cluster structure and on the geometry of the corresponding M sets. 

\subsection{Classifying synchronization in 3-dimensional CQNs}

\begin{figure}[h!]
\begin{center}
\includegraphics[width=\textwidth]{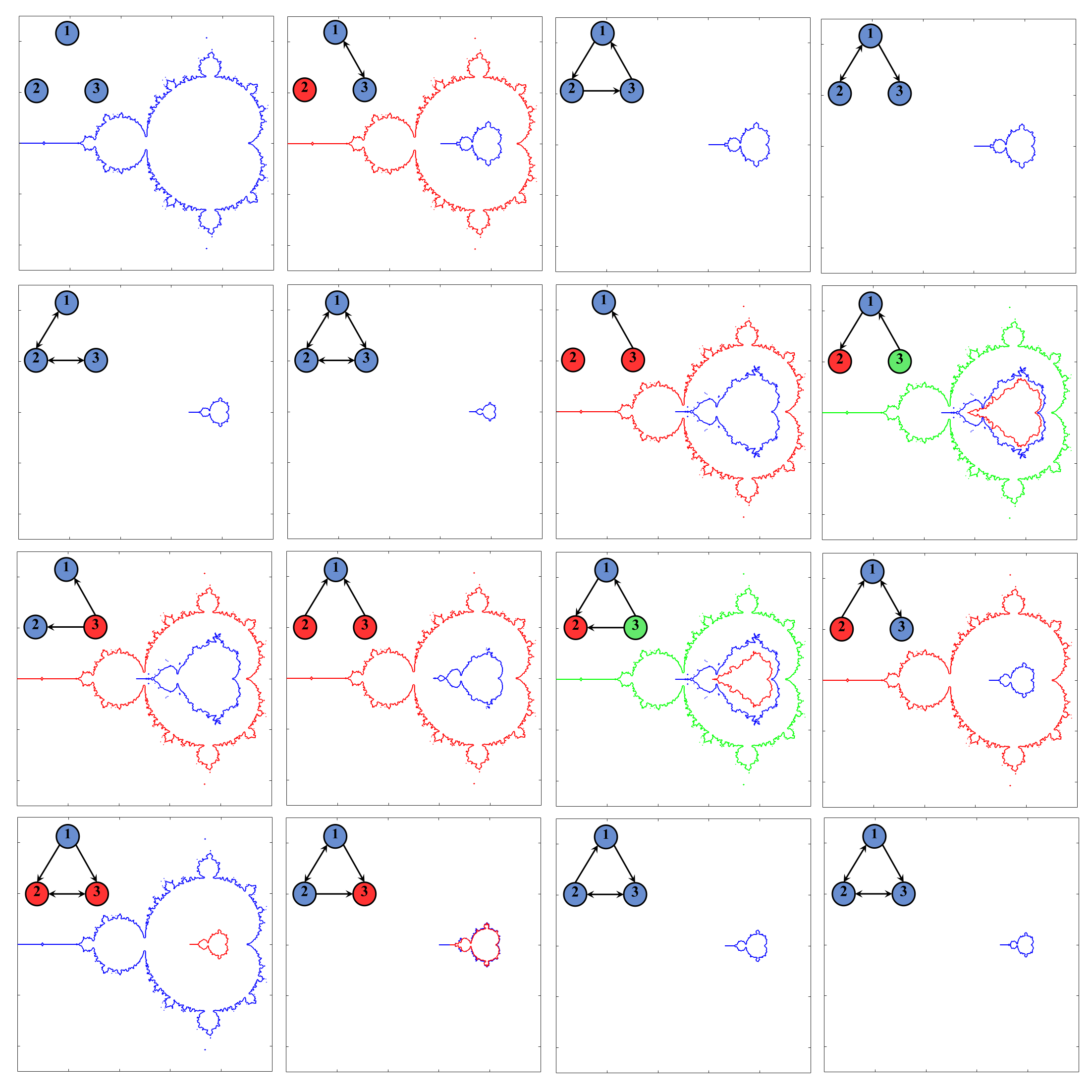}
\end{center}
\caption{\emph{\small {\bf Clustering behavior for all 16 normalized oriented networks, with three nodes.} Each node has a self-loop (not shown in the interest of space). Nodes that belong to the same M-cluster are shown in the same color.}}
\label{all3D}
\end{figure}

Figure~\ref{all3D} illustrates the shapes of the node-wise M sets and their clustering behavior for all possible oriented networks of three nodes, all three of which have self-loops, and such that all active edge weights = 1. Via symmetries, there are 16 such networks. Each figure panel corresponds to one of these networks (shown in the top left corner without the self loops, to avoid cluttering). We will refer to these panels in sequential order 1-16, from left to right and then top to bottom.

For each network, the system's M-set may synchronize into one, two or three M-clusters. We used different colors (blue, red and green) to distinguish the M-sets in different clusters, as well as to label the network nodes that correspond to each individual cluster. The first node $z_1$ is always labeled as ``cluster 1'' and marked in blue, together with all nodes that synchronize with it. The first node (in increasing order of the indices) that is not part of cluster 1 defines ``cluster 2'' and is marked in red, together with all the nodes that synchronize with it. If there is any node left that is not already in clusters 1 or 2, it automatically forms cluster 3, and is marked in green. Seven of the 16 cases lead to a single M-cluster, seven of them lead to two M-clusters, and the remaining two lead to the formation of three M-clusters. In some cases, the mechanisms behind the clustering behavior are intuitively clear; in other cases, the explanations are more subtle, or even counter-intuitive. 

For example notice, rather trivially, that symmetries between nodes lead to synchronization of those nodes into the same cluster. For example, in panels 1, 3 and 6, all nodes are identically coupled in the network, hence they have identical dynamics and they all synchronize into one cluster. It was pointed out in prior studies that higher node degrees lead to smaller network M sets, which is also reflected here. In panels 2, 9, 10, 13, it is also easy to notice why two of the nodes have identical wiring, hence they synchronize, while the behavior of the third node is different, and defines a second cluster. It is still somewhat intuitive why there are three different clusters for a network like that in panel 8 (representing a feed-forward network with self loops): node $z_3$ is independent on the other nodes dynamics, hence its M set is the traditional Mandelbrot set; $z_1$ feeds directly on $z_3$, and the integrated self and external input lead to a smaller M set for node $z_1$; in turn, $z_1$ projects onto $z_2$, leading to an M set for $z_2$ that is a subset of that corresponding to $z_1$ (hence the three clusters).

In this simplified scheme that analyzes only three-node interactions, it is therefore more tractable to tease apart which aspects of the network wiring lead to which which clustering scheme. Once the cluster formation is justified in each case, one interesting point of discussion is to track how changes in networking trigger synchronization changes. For example, if the connection from $z_1$ to $z_2$ in panel 8 is used instead to connect $z_3$ to $z_2$ (panel 9), then the node $z_2$ will no longer form its own cluster (with smaller M set than that of $z_1$), but instead it will join $z_1$ in its cluster, with no modifications to its M set. Interestingly, the same exact effect is produced on the M sets if the edge is simply removed altogether (panel 7). If, on the other hand, one restarts with the network in panel 8 and adds feedback responses along both inter-node inputs, all nodes synchronize to a new, smaller M set (panel 5). If one adds feedback just along one of these two edges, the network will switch to a two cluster structure (panel 12).

\subsection{Comparison with the Kuramoto model}
\label{Kuramoto}

In Section~\ref{effects}, we illustrated a simple example of how perturbations in connectivity affect synchronization patterns in a Kuramoto network. Here, we focus on describing similarities and differences between synchronization in Kuramoto networks and equi-M set synchronization in CQN networks. We discuss possible advantages of using equi-M set synchronization as a classification tool over the standard synchronization techniques. 

Instead of embracing standard synchronization in temporal activity, in CQNs we use equi M sets to describe synchronization of the nodes' critical behavior (bounded versus unbounded) in the complex parameter plane. For simple iterations of polynomials, it is well known that the critical orbits encode information about \emph{all} accessible orbit behaviors. For example, for traditional iterations of quadratic functions in the family $f_c(z) = z^2+c$, the Fatou-Julia theorem states that, for values of $c$ in the Mandelbrot set (i.e., the iteration of the critical point $\xi_0=0$ is bounded) all other initial conditions with bounded orbits (``prisoners'') form a connected set around $\xi_0$. In previous work, we have started exploring to what extent such a characterization remains true in the case of networks~\cite{radulescu2019asymptotic}. In the meantime, equi-M sets represent a way to visualize the sets of parameters for which different CQN nodes have different behavior.

\begin{figure}[h!]
\begin{center}
\includegraphics[width=0.8\textwidth]{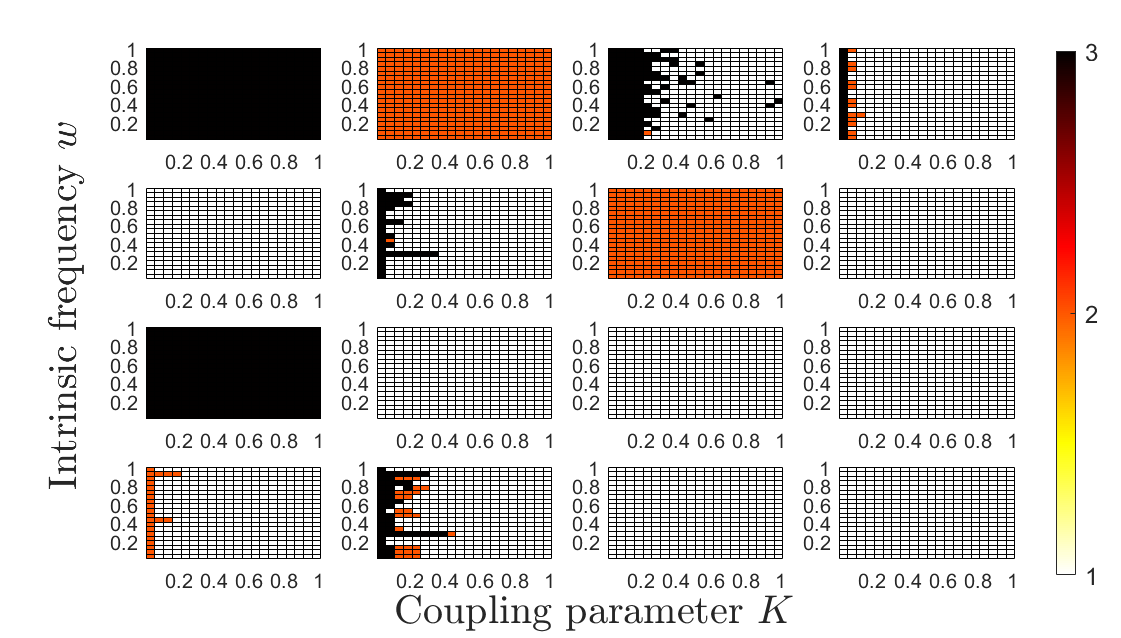}
\end{center}
\caption{\emph{\small {\bf Dependence of clustering parameters}, for all 3-dimensional oriented networks with self loops. For each network, the number of synchronization clusters was computed by a simple Matlab algorithm, for different parameter pairs $(K,w)$. Each panel represents one network (in the same order as in Figure~\ref{all3D}). Values of $K$ and $w$ in the interval [0,1] were sampled at a step of 0.2. In the $(K,w)$ parameter plane, the colors represent the number of clusters formed in the corresponding network, at the specified parameter values.}}
\label{param}
\end{figure}

\begin{figure}[h!]
\begin{center}
\includegraphics[width=0.8\textwidth]{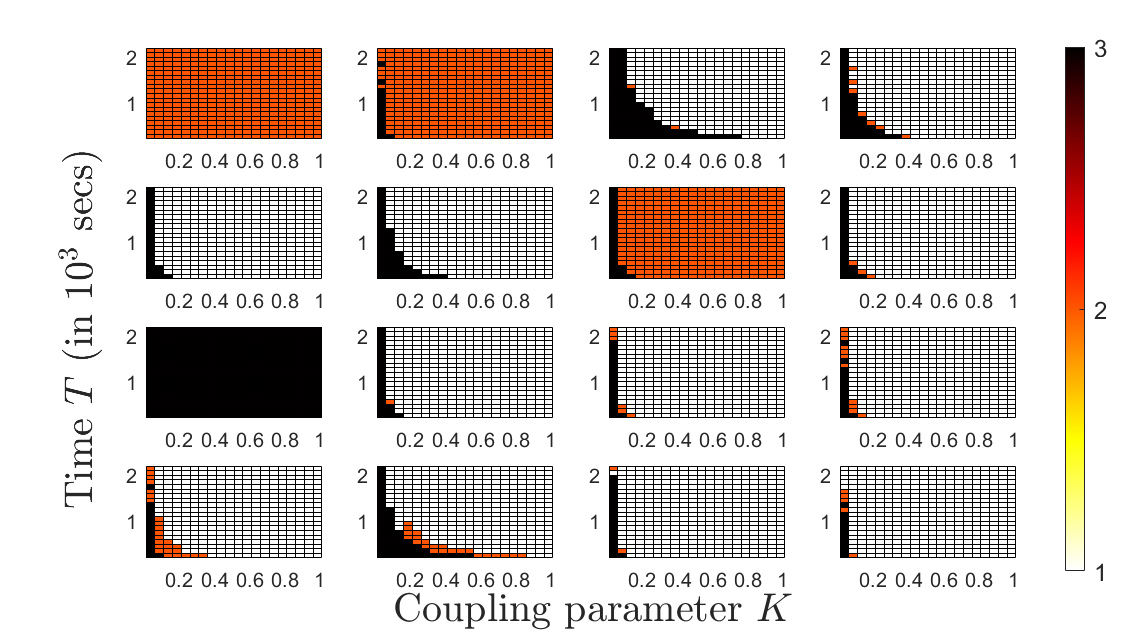}
\end{center}
\caption{\emph{\small {\bf Dependence of clustering on coupling and timing}, for all 3-dimensional oriented networks with self loops. For each network, the number of synchronization clusters was computed for different values of the connectivity parameter $K$ (in [0,1], sampled at a 0.2 size step), and at different times $T$ from 100 to 2000 secs (sampled in increments of 100 secs). In the $(K,T)$ plane, the colors represent the number of clusters formed in the corresponding network for the specified $K$ value, at the end of the time interval $[0,T]$ (with color coding as shown in the color bar). For all simulations, the intrinsic frequency $w$ was fixed to $w=1$.}}
\label{hybrid}
\end{figure}

Just like for CQNs with identical nodes, where the node dynamics are specified by the complex parameter $c = (\text{Re}(c), \text{Im}(c))$, the Kuramoto network with identical nodes~\eqref{Kuramoto_net} is controlled via a two-dimensional parameter plane $(w,K)$. By analogy with the classification of M-clusters in 3-dimensional CQNs, shown in Figure~\ref{all3D}, in Figure~\ref{param} we illustrate the parameter plane classification of phase coherence for the corresponding 3-dimensional networks of Kuramoto oscillators. More precisely, for each of the same 16 adjacency matrices described in the case of CQNs (Figure~\ref{all3D}), for each parameter pair in the region $(w,K) \in [0.1,1.1] \times [0.1,1.1]$, we considered the orbit initiated at balanced initial conditions $(2\pi/3,4\pi/3,2\pi)$, and we estimated the number of synchronization clusters after $T=500$ seconds. Notice that, as previously discussed, these results may change when $T$ is increased, in ways which depend on the parameter values and on the network itself (as further illustrated in Figure~\ref{hybrid}).

This all suggests that our M-synchronization approach is a more efficient tool when evaluating how emerging dynamic behavior depends on the network structure. For example, there are eight identical ``phase coherence'' parameter planes in the Kuramoto model simulations, all with $R=1$ (identifying a unique synchronization cluster throughout the region of interest in the $(w,K)$ plane). These correspond to widely different connectivity matrices, with different graph theoretical properties, differences which may be relevant to brain function and behavior, and which are not picked up by the synchronization patterns in a network model with Kuramoto node dynamics. As a very particular example consider networks 8 and 11 in the grid in Figure~\ref{all3D} (read top-down, then left to right). They represent a two-link motif and a triangular motif, respectively; the difference between the strengths of these motifs are believed to be a potentially important factor contributing to modulating brain activity and behavior. The Kuramoto model does not distinguish well between the patterns generated in these two networks. In turn, the M-synchronization technique in CQNs found an interesting difference. Even though the system has three M-clusters in both situations, the equi-M set corresponding to the node with the additional input in the triangular motif is small than in the two-link motif (a subset). In empirical work, we identified the size of the equi-M set to be a marker for gender and other physiological differences. It is very promising that such a marker can be tracked down to the shape of the network when using CQNs. Among other factors, one advantage of CQNs consists of being able to observe the behavior of each identical node (critically bounded or not bounded) independently from the others, without ignoring the connectivity web between them. Moreover, the geometric properties of equi-M sets represent as many ways to assess the effects of the underlying network on dynamics -- a set of measures which is substantially more complex and subtle than the phase coherence tools in Kuramoto networks. While the two models will always describe different types of dynamics, the M-synchronization appears to be a model that is better equipped to answer questions on the relationship between network architecture and the emerging coupled dynamics.

\subsection{M-synchronization in tractography-based networks}
\label{tractography}

To bridge this concept with its potential practical applications to neuroscience, we use human tractography data to illustrate a practical  example where M set synchronization can be used to understand the role of weak connections within the brain connectome.\\

\noindent \emph{\textbf{Data description and processing.}} Our analysis is based on tractography-derived, neural connectivity data obtained from the S1200 release of the Human Connectome Project~\cite{van2013wu}. The data set and the structural network construction are described in detail in Appendix C. In resulting structural networks provide us for the mathematical analysis with non-oriented, unsigned, normalized connectomes for 197 subjects. For each subject, the brain was partitioned into 116 areas (which we use a our network nodes), and normalized connection weights were obtained between each node pair (as described in Appendix C). Since these connectomes are tractography-based only, they do not capture either the directional aspect or the signature (excitatory vs. inhibitory) of the information transfer between node pairs. The corresponding matrices are therefore symmetric matrices with positive entries.

\begin{figure}[h!]
\begin{center}
\includegraphics[width=\textwidth]{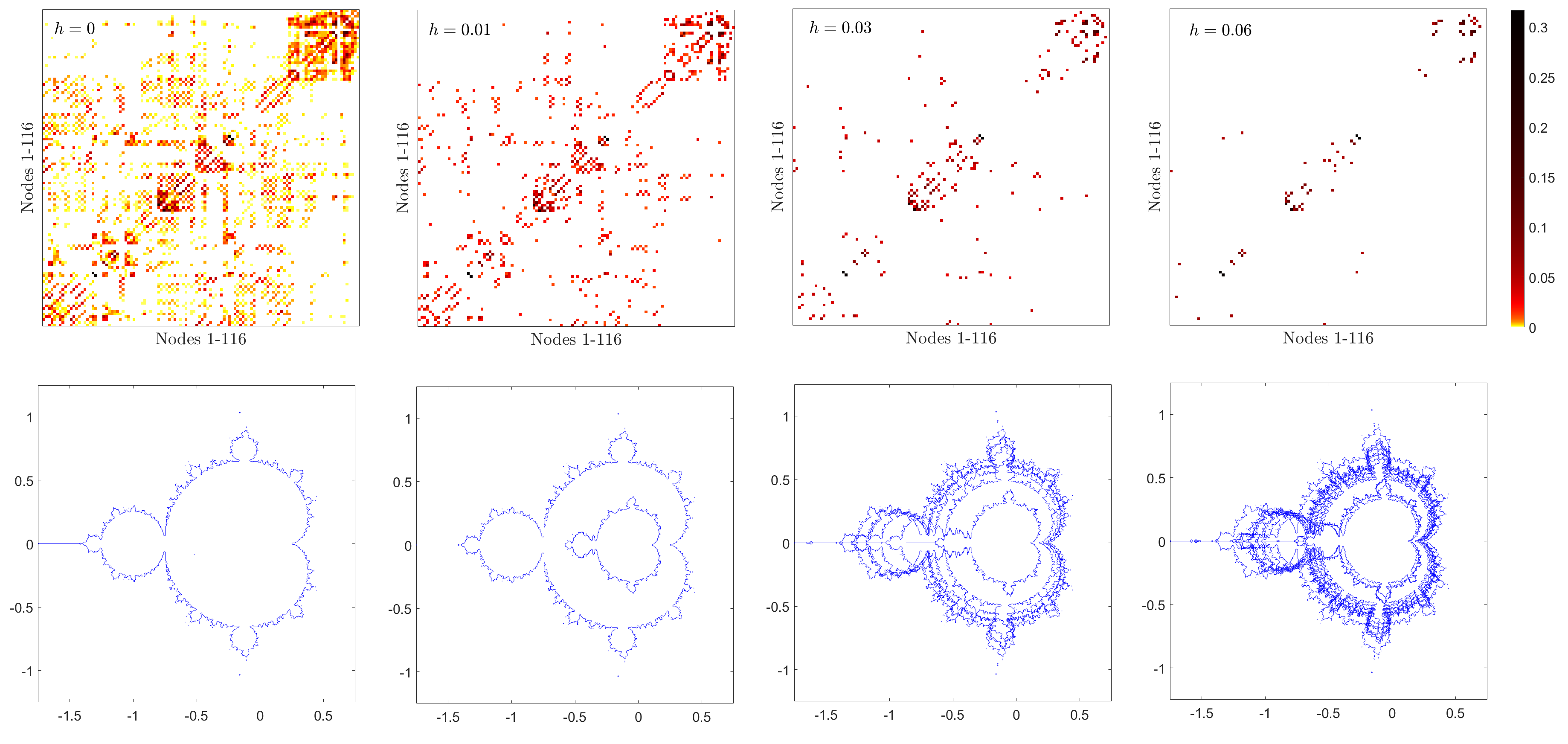}
\end{center}
\caption{\emph{\small {\bf Example of nodes de-synchronizing when weak connections are deleted.} From left to right, the top panels show respectively the original connectome, then the thresholded connectomes for $h=0.01$, $h=0.03$ and $h=0.05$. The bottom panels show the corresponding synchronization clusters for each case, increasing in number as $h$ is increased: one cluster (fully synchronized nodes) for the original connectome, two clusters for $h=0.01$, six clusters for $h=0.03$ and twelve clusters for $h=0.05$.}}
\label{contours_th}
\end{figure}

For this mathematical exploration, we chose to ``enhance'' the connectome of each subject by adding the identity matrix to the tractography-based connectivity matrix (i.e., we filled the first diagonal with unit entries).  We did so for two reasons. First (a mathematical reason): the inputs coming from other nodes are collectively very weak, preventing many receiving nodes from escaping to infinity, and thus leading to many trivial node-wise M sets (equivalent to the whole complex plane). Second (a modeling reason): this also reflects the corresponding situation in real brain networks, in which nodes self-regulate, even though these ``self-loops'' cannot be captured by tractography data.

From each of these ``enhanced'' matrices, we used Matlab to compute the node-wise M sets for all the 116 nodes (using 50 iterations of the network, with spacial resolution $200 \times 200$ pixels in the $c$-plane). We identified two nodes as being M-synchronized based on their M-sets occupying the same locus (via a maximum error $\epsilon =10$ pixels). Using the original connections, all the nodes were synchronized (i.e., showing only one M set cluster). However, when progressively pruning out the weakest connections, we observed the nodes subsequently de-synchronising (see Figure~\ref{contours_th} for an illustration). This is not surprising, since a low level of cross-talk between nodes is expected to contribute to their synchronization, hence eliminating some of that cross-talk would lead to a more diverse synchronization profile. However, as one would suspect, this increasing relationship between the number of clusters versus the threshold value cannot possibly extend universally, since eventually some nodes will be stripped of all interconnections, and, being driven solely by the unit self-input, will synchronize to the traditional Mandelbrot set.

To more consistently follow the effect of eliminating weak connections on synchronization, we used a moving threshold $h$ which we gradually increased from $h=0.01$ to $h=0.1$ in increments of $\Delta h = 0.005$. For each value of $h$, we then counted the number of different node-wise equi-M 
clusters, when deleting all connections weaker than the set threshold $h$. To give an idea of the effect of such thresholding on the connectome, let us notice that the low level cross-talk is prevalent. While the maximum strength connection among individuals ranges from $0.15-1.04$, the mean weight of active connections among individuals ranges between $0.006-0.012$. Hence when we delete, for example, all connections weaker than $h=0.01$ , only the strongest 16-32\% connections are retained (with different individuals placing differently within this range). When deleting all connections weaker than $h=0.03$, we are left with only the top 4-11\% connections; and when deleting all connections weaker than $h=0.05$, only the strongest 1-5\% connection are left.

\begin{figure}[h!]
\begin{center}
\includegraphics[width=0.7\textwidth]{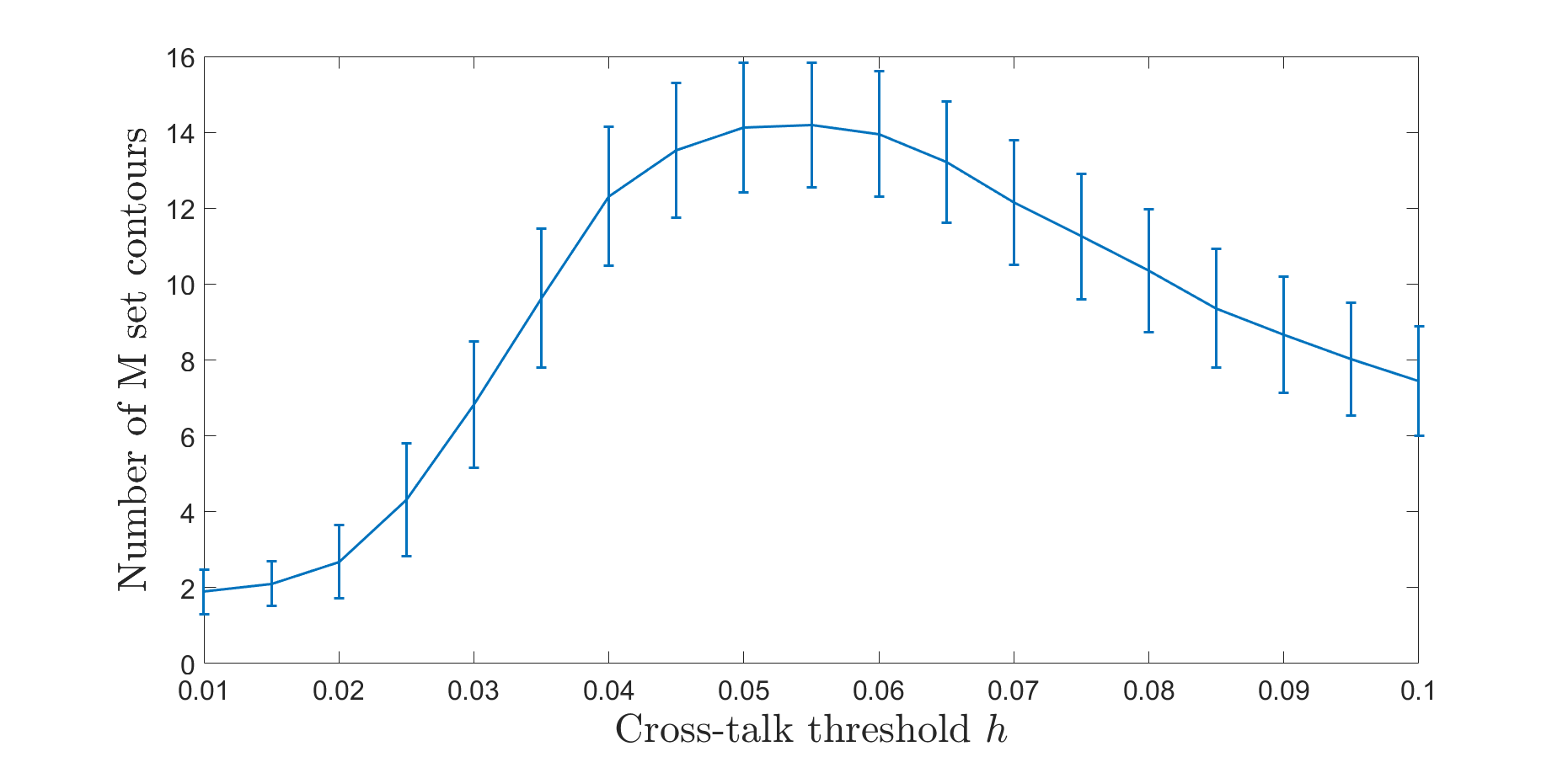}
\end{center}
\caption{\emph{\small {\bf Group statistics showing synchronization level relative to threshold value.} The curve shows the mean number of synchronization clusters (over the whole subject population), as it evolves when the threshold value $h$ is increased from $h=0.01$ to $h=0.1$. The error bars show the variability within the population. The number of contours was estimated (from contours in 200x200 spatial resolution), using an in house Matlab counting code.}}
\label{mean_curve}
\end{figure}

\begin{figure}[h!]
\begin{center}
\includegraphics[width=0.35\textwidth]{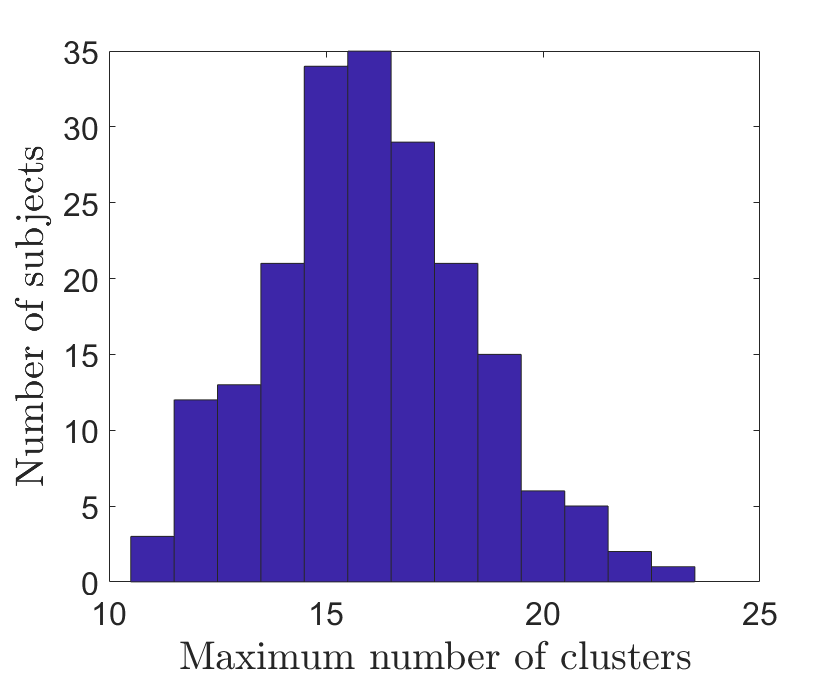}
\quad
\includegraphics[width=0.35\textwidth]{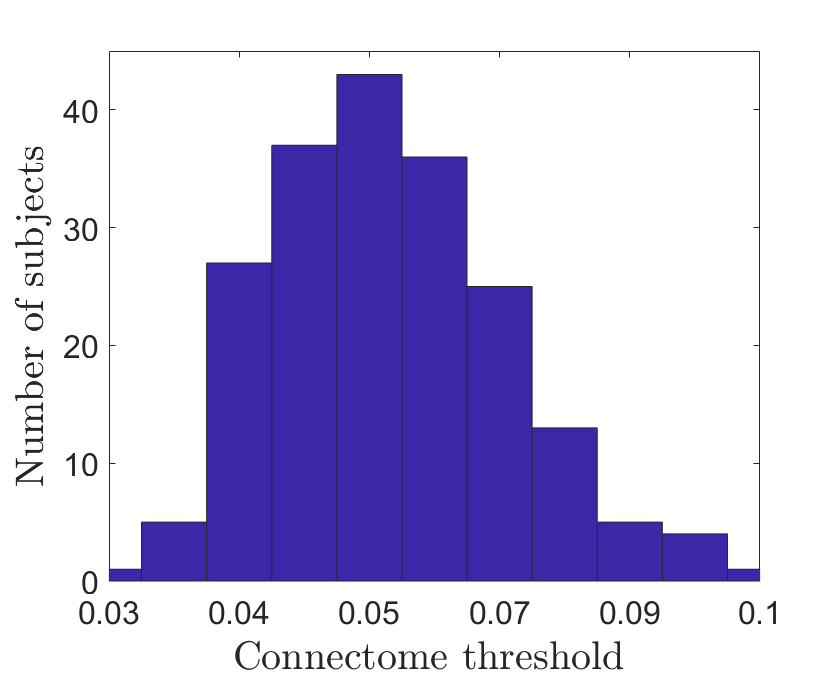}
\end{center}
\caption{\emph{\small {\bf Group statistics showing synchronization level and dependence on threshold value.} {\bf A.} Histogram showing the distribution of the maximum number of clusters across all subjects. {\bf B.} Histogram showing the distribution of the threshold value where the maximum number of clusters occurs, across all subjects.}}
\label{histograms}
\end{figure}

With this in mind, we see the progression of de-synchronization as the threshold $h$ is being increased. Figure~\ref{mean_curve} shows the mean number of distinct M clusters across the subject population, with error bars showing between-subject variability, and how this mean evolves with respect to the increasing threshold $h$. Notice that the maximum desynchronization is reflected as 11-23 clusters (with variability between individuals, as shown in the histogram in Figure~\ref{histograms}a), and these maxima are achieved for more than half of the subjects within the interval $h=0.05-0.06$ (the variability between subjects is shown in the histogram in Figure~\ref{histograms}b).

First, the maximum desyncrhonization is consistently achieved when most of the cross-talk between nodes has been eliminated, and only the strongest 1-5\% of  connections are left. While this may be initially counter-intuitive, it in fact relates to existing knowledge of information processing in dynamic networks. The idea that sparse networks are useful for brain processing, and may in fact outperform denser networks in certain contexts is not new~\cite{litwin2017optimal}. In sparse networks, the contributions of individual nodes are less centralized, allowing more flexibility of the ensemble response. In our context, this reflected in the fact that sparser networks exhibit a larger number of clusters, while denser networks are more likely to lead to node synchronization, and diminishing the within-network variability that may be required for accomplishing certain brain functions.

The low-level cross-talk between brain areas may allow the brain to swiftly switch between synchronization modes, in order to produce a desired functional outcome. Our results suggest that the physical mechanism used to sustain this possibility may involve maintaining a multitude of weak connections, which can be used as needed for active information transfer, based on context and desired outcome. Broad and quick decisions on how much cross-talk to actively involve in a certain process is consistent with our knowledge of both brain plasticity and dynamics. This mechanism may be a cost reasonable choice for the system to efficiently accomplish synchronization switching, since weak connections are more inexpensive to maintain, and easier to sacrifice. To determine which of the anatomical connections between nodes are actually in use at each moment, would require an imaging modality that reflects functional connectivity. 


\section{Discussion}

\subsection{General comments on synchronization and clustering}

In this paper, we explored the potential of implementing a synchronization concept driven from applications in network neuroscience, to the theoretical context of networks with complex quadratic nodes. The aim of this is to facilitate analytical approaches to the complex phenomena in nonlinear networks of coupled neural oscillators, by drawing analogies with canonical networks with more tractable dynamics (such as CQNs).

We determined that network nodes separate into a finite number of clusters, based on the criterion that members of each cluster have identical M sets. The cluster number and grouping are primarily determined by the distribution of edges in the network, and the geometry of the M sets in each cluster is further tuned by the connectivity patterns and weights. 
We found that the network equi-M set (defined as the intersection of the node-wise M sets) may or may not be a nested intersection, depending on the network architecture.  

One significant limitation relates to interpreting some of the illustrations based on numerical computations. That is because the existence of escape radius has been determined for a few network types, such as feed forward networks with self loops and dominated networks, but is not known in general. This requires additional caution when interpreting some of the illustrations, which are based on a finite number of iterations; the potential exists for the orbits to remain bounded after escaping the large disc set in the simulation code as the escape domain. Establishing and calculating escape radii for more network classes is one of the primary goals of our future work, and will increase the confidence when interpreting simulations.

\subsection{Connections with other models and applications}

Part of our goal when studying CQNs consists of understanding to what extent methods from traditional, single map discrete dynamical systems can be extended to the context of networks. The other aspect of this approach is that of understanding how behaviors in CQN theoretical networks may be universal, or similar phenomenologically to behaviors of networks of interest in the natural sciences. Translating phenomena such as ``node synchronization'' and ``clustering'' into the framework of CQNs is a significant step in this latter direction.

The idea that network connectivity governs the nodes' synchronization and clustering is a central theme in computational neuroscience. For example, modeling work in basal ganglia networks found that different types of connectivity patterns lead to different synchronization clusters and spiking rhythms~\cite{terman2002activity}. Other studies show that even in networks with fixed architecture (all-to-all inhibitory reticulate thalamic networks) small perturbations in synaptic conductance can trigger restructuring of the nodes into different numbers of synchronized clusters~\cite{golomb1994clustering}. While our model is very different from the ones above, both in the type of dynamics (discrete versus continuous time), and in the definition of synchronization, our results capture efficiently dependencies of clustering on local and global perturbations to the network structure and weights. In fact, a simple comparison suggested that our CQN model outperforms Kuramoto networks in terms of capturing how network perturbations affect synchronization patterns.

Our exploration of the brain tractography data emphasized the idea that weak background cross-talk between various parts of the brain plays an extremely important role in syncrhonizing dynamics. This has been observed before in networks (e.g., opinion formation in social networks), but has never been fully explained. A complex network like the brain has to function within an optimal range, where it can swiftly adapt to ``synchronize'' or ``de-synchronize,'' and may be able to do so by using small changes to the level of cross-talk. This study shows how our abstract mathematical approach can be used to capture the link between weak connections and brain synchronization patterns. 

A future direction, tying into applications and personalized assessments,  is to further study between-subject differences in synchronization profile and threshold effects. As Figures~\ref{mean_curve} and~\ref{histograms} show, there is variability between the number of synchronization clusters for different subjects for any given threshold. Moreover, the maximum number of clusters differs in value, and is achieved at different thresholds between individuals. This likely reflects different efficiency levels between individuals when switching synchornization regimes, and may be significant to which dynamic patterns are more likely to occur in each subject. 

More generally: as we are slowly starting to learn more about network dynamics, reservoir pools, deep learning and artificial intelligence, a lot of important questions in these fields yet remain unanswered by the traditional methods and techniques associated with dynamic networks. Our study illustrates how complex dynamics and our extension of Julia-Mandelbrot theory can be used to contextualize and address some of these questions in a tractable framework.

\section*{Acknowledgements}

The project received support from the Simons Foundation (R\v{a}dulescu, \#523763) and from the SUNY New Paltz AC$^2$ program.

Data were provided [in part] by the Human Connectome Project, WU-Minn Consortium (Principal Investigators: David Van Essen and Kamil Ugurbil; 1U54MH091657) funded by the 16 NIH Institutes and Centers that support the NIH Blueprint for Neuroscience Research; and by the McDonnell Center for Systems Neuroscience at Washington University.

\section*{Appendix A}

A simulation based on 1000 iterates supports the fact that, for $c=-0.2+0.06$, the node $z_{13} \to -0.4601+0.0889i$ (hence remains bounded), while the imaginary part of the component $z_{12}$ blows up to infinity (hence the node escapes). On the other hand, for $c=-0.23+0.05i$, the component $z_{12}$ approaches a periodic two oscillation between $-0.4825 + 0.1075i$ and $-0.0175 - 0.1075i$, while the imaginary part of $z_{13}$ escapes. 

\section*{Appendix B}

The matrices corresponding to the networks in the three panels of Figure~\ref{20dim_M_clusters} are described below, in order of the panels:

\begin{tiny}
$A_a = \left ( \begin{array}{cccccccccccccccccccc}
   1 &  0 &  0 &  0 &  0 &  0 &  1 &  0 &  0 &  0 &  0 &  0 &  0 &  1 &  1 &  0 &  0 &  0 &  1 &  0\\
   0 &  1 &  0 &  0 &  0 &  0 &  0 &  1 &  1 &  0 &  0 &  0 &  0 &  0 &  0 &  0 &  0 &  0 &  0 &  0\\
   0 &  0 &  1 &  0 &  0 &  0 &  0 &  0 &  0 &  0 &  1 &  0 &  0 &  0 &  0 &  1 &  0 &  0 &  1 &  0\\
   0 &  0 &  1 &  1 &  0 &  0 &  0 &  0 &  0 &  0 &  0 &  0 &  0 &  0 &  0 &  0 &  0 &  0 &  0 &  0\\
   0 &  0 &  0 &  0 &  1 &  0 &  0 &  0 &  0 &  1 &  0 &  0 &  1 &  0 &  0 &  0 &  0 &  0 &  0 &  0\\
   1 &  0 &  0 &  0 &  0 &  1 &  0 &  1 &  0 &  0 &  0 &  0 &  0 &  0 &  0 &  0 &  0 &  1 &  0 &  0\\
   0 &  0 &  0 &  0 &  0 &  1 &  1 &  0 &  0 &  0 &  0 &  0 &  0 &  0 &  0 &  0 &  0 &  0 &  0 &  0\\
   0 &  0 &  1 &  0 &  0 &  0 &  0 &  1 &  0 &  0 &  0 &  0 &  0 &  0 &  0 &  0 &  1 &  0 &  0 &  0\\
   0 &  0 &  0 &  1 &  0 &  1 &  0 &  0 &  1 &  0 &  0 &  0 &  0 &  0 &  0 &  0 &  0 &  0 &  0 &  0\\
   0 &  1 &  0 &  0 &  0 &  0 &  0 &  0 &  0 &  1 &  1 &  1 &  0 &  0 &  0 &  0 &  0 &  0 &  0 &  0\\
   0 &  0 &  0 &  0 &  0 &  0 &  1 &  1 &  1 &  0 &  1 &  0 &  0 &  0 &  0 &  0 &  0 &  0 &  0 &  0\\
   0 &  0 &  0 &  1 &  0 &  0 &  0 &  0 &  0 &  0 &  1 &  1 &  0 &  0 &  0 &  0 &  1 &  0 &  0 &  0\\
   0 &  0 &  0 &  0 &  0 &  0 &  0 &  0 &  0 &  0 &  0 &  0 &  1 &  0 &  0 &  0 &  0 &  0 &  0 &  0\\
   0 &  0 &  0 &  0 &  0 &  0 &  0 &  0 &  1 &  0 &  1 &  0 &  0 &  1 &  0 &  0 &  0 &  0 &  0 &  0\\
   0 &  0 &  1 &  0 &  0 &  0 &  0 &  0 &  0 &  0 &  0 &  0 &  0 &  1 &  1 &  0 &  0 &  0 &  0 &  0\\
   0 &  0 &  0 &  0 &  0 &  0 &  0 &  0 &  0 &  0 &  0 &  0 &  0 &  0 &  0 &  1 &  0 &  0 &  0 &  0\\
   0 &  0 &  0 &  0 &  0 &  0 &  0 &  0 &  0 &  0 &  0 &  0 &  0 &  0 &  0 &  1 &  1 &  0 &  0 &  0\\
   0 &  0 &  0 &  0 &  0 &  0 &  0 &  0 &  1 &  1 &  0 &  0 &  0 &  0 &  0 &  0 &  0 &  1 &  0 &  0\\
   0 &  1 &  0 &  0 &  0 &  0 &  0 &  0 &  1 &  0 &  0 &  0 &  0 &  0 &  1 &  0 &  0 &  0 &  1 &  0\\
   0 &  0 &  0 &  0 &  0 &  0 &  0 &  0 &  0 &  0 &  0 &  0 &  0 &  0 &  0 &  0 &  0 &  0 &  0 &  1
\end{array}
\right)$

\vspace{3mm}
$A_b = \left ( \begin{array}{cccccccccccccccccccc}
   1 &  0 &  0 &  0 &  0 &  0 &  1 &  0 &  0 &  0 &  0 &  0 &  1 &  0 &  0 &  0 &  1 &  0 &  0 &  0\\
   0 &  1 &  0 &  0 &  0 &  0 &  0 &  1 &  0 &  0 &  0 &  0 &  0 &  0 &  0 &  0 &  0 &  0 &  0 &  0\\
   0 &  0 &  1 &  1 &  0 &  0 &  1 &  0 &  0 &  0 &  0 &  0 &  0 &  0 &  1 &  0 &  0 &  0 &  0 &  0\\
   0 &  1 &  0 &  1 &  1 &  0 &  0 &  0 &  0 &  0 &  0 &  0 &  0 &  0 &  0 &  0 &  0 &  1 &  0 &  0\\
   0 &  0 &  0 &  1 &  1 &  0 &  0 &  0 &  0 &  0 &  0 &  0 &  0 &  0 &  0 &  0 &  0 &  0 &  0 &  0\\
   0 &  0 &  0 &  0 &  0 &  1 &  0 &  0 &  0 &  0 &  0 &  1 &  0 &  0 &  0 &  0 &  0 &  0 &  0 &  0\\
   0 &  0 &  0 &  0 &  0 &  0 &  1 &  0 &  0 &  0 &  0 &  1 &  0 &  0 &  0 &  0 &  0 &  0 &  0 &  0\\
   0 &  0 &  0 &  0 &  0 &  0 &  0 &  1 &  0 &  0 &  0 &  0 &  0 &  0 &  0 &  0 &  0 &  0 &  0 &  0\\
   0 &  0 &  0 &  0 &  0 &  0 &  0 &  0 &  1 &  0 &  0 &  0 &  0 &  0 &  0 &  0 &  0 &  0 &  0 &  0\\
   0 &  0 &  0 &  0 &  0 &  0 &  0 &  0 &  0 &  1 &  0 &  0 &  0 &  0 &  0 &  0 &  1 &  0 &  0 &  0\\
   0 &  0 &  0 &  0 &  0 &  0 &  0 &  0 &  0 &  1 &  1 &  0 &  0 &  0 &  0 &  0 &  0 &  1 &  0 &  0\\
   0 &  0 &  0 &  0 &  0 &  0 &  0 &  0 &  0 &  0 &  0 &  1 &  0 &  0 &  0 &  0 &  0 &  0 &  1 &  0\\
   0 &  1 &  0 &  0 &  0 &  0 &  0 &  1 &  0 &  0 &  0 &  0 &  1 &  0 &  0 &  0 &  0 &  0 &  0 &  1\\
   1 &  0 &  0 &  0 &  1 &  0 &  0 &  0 &  0 &  0 &  0 &  0 &  0 &  1 &  0 &  0 &  0 &  1 &  0 &  0\\
   0 &  0 &  0 &  0 &  0 &  0 &  0 &  0 &  0 &  0 &  0 &  1 &  0 &  1 &  1 &  0 &  0 &  0 &  0 &  0\\
   0 &  0 &  0 &  0 &  0 &  0 &  0 &  0 &  1 &  0 &  1 &  0 &  0 &  0 &  1 &  1 &  0 &  0 &  0 &  0\\
   0 &  0 &  0 &  0 &  0 &  0 &  0 &  0 &  0 &  0 &  0 &  0 &  0 &  0 &  0 &  0 &  1 &  0 &  0 &  0\\
   1 &  0 &  0 &  0 &  0 &  0 &  0 &  0 &  0 &  0 &  0 &  0 &  0 &  0 &  0 &  0 &  0 &  1 &  0 &  0\\
   0 &  0 &  0 &  0 &  0 &  0 &  0 &  0 &  0 &  0 &  0 &  1 &  0 &  0 &  0 &  0 &  0 &  0 &  1 &  0\\
   0 &  0 &  0 &  0 &  0 &  0 &  0 &  0 &  0 &  0 &  0 &  0 &  0 &  0 &  0 &  0 &  0 &  0 &  0 &  1

\end{array}
\right)$

\vspace{3mm}
$A_c = \left ( \begin{array}{cccccccccccccccccccc}
   1 &  0 &  0 &  1 &  0 &  0 &  0 &  0 &  0 &  1 &  1 &  1 &  0 &  0 &  0 &  0 &  0 &  0 &  0 &  0\\
   1 &  1 &  0 &  1 &  0 &  0 &  0 &  0 &  1 &  1 &  0 &  0 &  0 &  0 &  0 &  0 &  0 &  0 &  0 &  1\\
   0 &  0 &  1 &  0 &  0 &  0 &  0 &  0 &  0 &  0 &  1 &  0 &  0 &  0 &  0 &  0 &  1 &  0 &  0 &  1\\
   0 &  1 &  1 &  1 &  0 &  0 &  0 &  0 &  1 &  0 &  0 &  0 &  0 &  0 &  0 &  0 &  1 &  0 &  0 &  0\\
   0 &  0 &  1 &  0 &  1 &  1 &  1 &  1 &  0 &  0 &  0 &  0 &  1 &  0 &  1 &  0 &  0 &  0 &  1 &  0\\
   1 &  1 &  0 &  0 &  0 &  1 &  0 &  0 &  1 &  1 &  0 &  0 &  0 &  0 &  0 &  0 &  0 &  1 &  1 &  0\\
   0 &  1 &  0 &  0 &  0 &  0 &  1 &  0 &  1 &  1 &  1 &  0 &  0 &  1 &  0 &  1 &  0 &  0 &  1 &  0\\
   1 &  0 &  0 &  0 &  0 &  0 &  1 &  1 &  0 &  0 &  0 &  0 &  0 &  0 &  0 &  0 &  1 &  0 &  0 &  0\\
   0 &  0 &  0 &  0 &  0 &  0 &  1 &  0 &  1 &  0 &  1 &  0 &  0 &  0 &  1 &  0 &  1 &  0 &  0 &  0\\
   0 &  1 &  0 &  1 &  0 &  0 &  0 &  0 &  0 &  1 &  1 &  0 &  0 &  1 &  0 &  0 &  0 &  0 &  0 &  1\\
   1 &  0 &  0 &  1 &  0 &  0 &  1 &  0 &  0 &  0 &  1 &  1 &  1 &  0 &  1 &  0 &  0 &  0 &  1 &  0\\
   1 &  1 &  1 &  0 &  0 &  0 &  0 &  0 &  1 &  1 &  0 &  1 &  0 &  1 &  0 &  0 &  1 &  0 &  1 &  0\\
   0 &  0 &  1 &  0 &  0 &  1 &  0 &  0 &  0 &  0 &  0 &  0 &  1 &  0 &  0 &  1 &  0 &  0 &  0 &  1\\
   0 &  0 &  1 &  0 &  0 &  1 &  0 &  0 &  0 &  0 &  0 &  0 &  0 &  1 &  0 &  0 &  0 &  1 &  1 &  0\\
   0 &  0 &  0 &  0 &  0 &  0 &  0 &  0 &  0 &  0 &  1 &  1 &  1 &  0 &  1 &  0 &  0 &  0 &  0 &  0\\
   1 &  0 &  1 &  0 &  0 &  0 &  0 &  0 &  0 &  1 &  1 &  0 &  0 &  0 &  1 &  1 &  0 &  0 &  1 &  1\\
   0 &  0 &  1 &  1 &  0 &  0 &  0 &  0 &  0 &  1 &  0 &  0 &  0 &  1 &  1 &  0 &  1 &  0 &  1 &  1\\
   0 &  1 &  0 &  1 &  0 &  0 &  0 &  1 &  1 &  1 &  0 &  0 &  0 &  0 &  0 &  1 &  0 &  1 &  0 &  0\\
   0 &  1 &  0 &  1 &  1 &  1 &  0 &  0 &  1 &  0 &  0 &  0 &  0 &  0 &  1 &  1 &  1 &  0 &  1 &  1\\
   1 &  0 &  0 &  1 &  1 &  1 &  1 &  0 &  1 &  0 &  0 &  0 &  0 &  0 &  0 &  1 &  0 &  0 &  1 &  1
\end{array}
\right)$

\end{tiny}

\section*{Appendix C: data and structural network construction}

The project released to the public domain extensive MRI, behavioral, and demographic data from a large cohort of individuals ($>1000$). For our own study, we considered a subject subgroup of $N=197$ individuals (90 males and 107 females, ages from 22 to over 36 years old). 

The original Structural Network Construction 
Preprocessed diffusion weighted images were used to construct structural connectomes for the subjects. Fiber tracking was done using DSI Studio with a modified FACT algorithm~\cite{yeh2013deterministic}. As a first step, data were reconstructed using generalized q-sampling imaging (GQI)~\cite{yeh2010generalized}. Diffusion weighted images were reconstructed in native space and the quantitative anisotropy (QA) for each voxel was computed. Fiber tracking was performed until 250,000 streamlines were reconstructed with angular threshold of 50o, step size of 1.25 mm, minimum length of 10mm, and maximum length of 400mm. 
Streamline counts were estimated for two parcellations schemes based on the AAL~\cite{tzourio2002automated} atlas version-1 containing 116. The AAL atlas was registered to the isotropic diffusion component (ISO) image, an output of GQI. Registering directly to the ISO image minimizes any registration issues that could arise by first registering to an individual’s T1w image. Atlas registration was conducted using FSL FLIRT~\cite{jenkinson2002improved} with default parameters. A template T1w image was registered to the subject-specific ISO image. The transformation matrix obtained from the registration was applied to all regions in the AAL atlas. In case two regions were registered to the same voxel, the voxel was assigned to the region with the highest probability. In order to fully sample the fiber orientations within a voxel, tracking was repeated 250x with initiation at a random sub-voxel position generating 250 connectivity matrices per subject.
For each subject, an undirected weighted structural connectivity matrix, A, was constructed from the connection strength based on the number of streamlines connecting two regions The final connectivity matrix included edges which were present in at least 10\% (or 25) of the 250 matrices. Connectivity matrix was normalized by dividing the number of streamlines between each two coupled regions, by the combined volumes of the two regions.

\bibliographystyle{unsrt}
\bibliography{references}

\end{document}